\documentclass{article}

\usepackage{arxiv}

\usepackage[toc,page]{appendix}

\usepackage{graphicx}
\usepackage{dcolumn}
\usepackage{bm}
\usepackage{appendix} 

\usepackage[mathlines]{lineno}

\usepackage[utf8]{inputenc}
\usepackage{gensymb} 
\usepackage{booktabs} 
\usepackage{subcaption} 
\usepackage{color} 
\usepackage{adjustbox}
\usepackage{amsmath}
\usepackage{mathtools, nccmath}
\usepackage{xparse}
\DeclarePairedDelimiterX{\set}[1]{\{}{\}}{\setargs{#1}}
\usepackage{listings}
\usepackage{enumitem}
\definecolor{codegreen}{rgb}{0,0.6,0}
\definecolor{codegray}{rgb}{0.5,0.5,0.5}
\definecolor{codepurple}{rgb}{0.58,0,0.82}
\definecolor{backcolour}{rgb}{0.95,0.95,0.92}
\usepackage{textcomp}

\usepackage{algpseudocode}
\usepackage{amsfonts}
\usepackage{lscape}
\usepackage{longtable}

\lstdefinestyle{mystyle}{
  backgroundcolor=\color{backcolour}, commentstyle=\color{codegreen},
  keywordstyle=\color{magenta},
  numberstyle=\tiny\color{codegray},
  stringstyle=\color{codepurple},
  basicstyle=\ttfamily\footnotesize,
  breakatwhitespace=false,         
  breaklines=true,                 
  captionpos=b,                    
  keepspaces=true,                 
  numbers=left,                    
  numbersep=5pt,                  
  showspaces=false,                
  showstringspaces=false,
  showtabs=false,                  
  tabsize=2
}

\usepackage[utf8]{inputenc}
\usepackage[english]{babel}
\usepackage{graphicx}
\usepackage{gensymb} 
\usepackage{booktabs} 
\usepackage{lipsum}

\usepackage[normalem]{ulem} 
\usepackage{bm}
\usepackage{bbm}
\usepackage{amssymb}
\usepackage{multirow}

\usepackage{url}
\usepackage[hidelinks]{hyperref}

\usepackage{algcompatible}
\usepackage{newfloat}
\DeclareFloatingEnvironment[
    fileext=loa,
    listname=List of Algorithms,
    name=ALGORITHM,
    placement=tbhp,
]{algorithm}
\newcommand{\beginsupplement}{%
        \setcounter{table}{0}
        \renewcommand{\thetable}{S\arabic{table}}%
        \setcounter{figure}{0}
        \renewcommand{\thefigure}{S\arabic{figure}}%
     }

\title{Proton arc therapy plan optimization with energy layer pre-selection driven by organ at risk sparing and delivery time}

\usepackage{authblk}

\newcommand{\mycomment}[1]{}

\author[1]{Sophie Wuyckens}
\author[2,3]{Guillaume Janssens}
\author[1]{Macarena Chocan Vera}
\author[4]{Johan Sundström}
\author[5]{Dario Di Perri}
\author[1,6,7]{Edmond Sterpin}
\author[1,2]{Kevin Souris}
\author[1]{John A. Lee}

\affil[1]{Université catholique de Louvain, Institut de Recherche Expérimentale et Clinique (IREC), Molecular Imaging, Radiotherapy and Oncology, Woluwe-Saint-Lambert, Belgium}
\affil[2]{Ion Beam Applications SA, Louvain-La-Neuve, Belgium}
\affil[3]{Université catholique de Louvain, ICTEAM, Louvain-La-Neuve, Belgium}
\affil[4]{RaySearch Laboratories AB, Stockholm, Sweden}
\affil[5]{Department of Radiation Oncology, Cliniques universitaires Saint-Luc, Brussels, Belgium}
\affil[6]{KULeuven, Department of Oncology, Laboratory of experimental radiotherapy, Leuven, Belgium}
\affil[7]{Particle Therapy Interuniversity Center Leuven - PARTICLE, Leuven, Belgium}

\renewcommand{\headeright}{Preprint}
\renewcommand{\undertitle}{Preprint}
\renewcommand{\shorttitle}{Flexi-Arc}

\begin{document}
\maketitle

\begin{abstract}
\textit{Objective.} As proton arc therapy (PAT) approaches clinical implementation, optimizing treatment plans for this innovative delivery modality remains challenging, especially in addressing arc delivery time. Existing algorithms for minimizing delivery time are either optimal but computationally demanding or fast but at the expense of sacrificing many degrees of freedom. In this study, we introduce a flexible method for pre-selecting energy layers (EL) in PAT treatment planning before the actual robust spot weight optimization. \\
\textit{Approach.}  Our EL pre-selection method employs metaheuristics to minimize a bi-objective function, considering a dynamic delivery time proxy and tumor geometrical coverage penalized as a function of selected organs-at-risk crossing. It is capable of parallelizing multiple instances of the problem. We evaluate the method using three different treatment sites, providing a comprehensive dosimetric analysis benchmarked against dynamic proton arc plans generated with early energy layer selection and spot assignment (ELSA) and IMPT plans in RayStation TPS.\\
\textit{Result.} The algorithm efficiently generates Pareto-optimal EL pre-selections in approximately 5 minutes. Subsequent PAT treatment plans derived from these selections and optimized within the TPS, demonstrate high-quality target coverage, achieving a high conformity index, and effective sparing of organs at risk. These plans meet clinical goals while achieving a 20 to 40\% reduction in delivery time compared to ELSA plans.  \\   
\textit{Significance.} The proposed algorithm offers speed and efficiency, producing high-quality PAT plans by placing proton arc sectors to efficiently reduce delivery time while maintaining good target coverage and healthy tissues sparing.
\end{abstract}


\section{Introduction}
Proton arc therapy (PAT) is an advanced modality of radiation therapy that dynamically delivers proton beams during gantry rotation. Compared to conventional techniques, such as intensity-modulated proton therapy (IMPT), multiple studies have shown that PAT could potentially offer superior dose distribution \cite{liu_lung_2021}, increase treatment robustness \cite{chang_feasibility_2020,liu_improve_2020}, enhance radiobiological effects \cite{bertolet_proton_2020,carabe_radiobiological_2020}, and simplify the treatment workflow \cite{liu_novel_2020}. However, the treatment planning optimization challenge associated with this novel modality demands sophisticated algorithms and innovative computational approaches to fully exploit its potential benefits while keeping realistic treatment times. 

In addition to the usual dose-based objectives, the PAT optimization problem introduces a new aspect, which is the minimization of the delivery time. The arc delivery time tightly depends on the number and sequence of energy layers (EL) and gantry angles. On the other hand, the PAT optimization space contains a much larger number of beam angles and EL candidates than conventional IMPT. Consequently, significant efforts have been made to minimize delivery time by reducing the number of EL and finding an energy sequence such that energy switch-up (SU) occurrences are minimized, as these transitions are the most time-consuming.

There are currently two major approaches to solving the PAT problem. First, one can explicitly include both dose and delivery time surrogate in the optimization function. For instance, Gu et al.~published a complex objective function, composed of several terms, each addressing a specific goal, such as EL sparsity and sequencing, during spot weight optimization \cite{gu_novel_2020}. Mathematically speaking, this integrated framework is the most rigorous and the solution it finds is likely optimal, harnessing the full set of degrees of freedom (DOFs). Zhang et al. and Wuyckens et al.~have also explored such methods \cite{zhang_energy_2022,wuyckens_treatment_2022,zhang_treatment_2023}. However, these methods share common drawbacks, including the need for new algorithms to handle non-differentiable objectives, tedious tuning of objective weights, and heavy computational demand. 

Secondly, there are heuristic-based methods that decouple the problem by selecting beforehand, or iteratively, the ELs and then performing a regular spot weight optimization with the selected set of ELs. Although these techniques eliminate part of the DOFs, they yield encouraging results and do so usually in a very reasonable computation time, making them more suitable for clinical integration. For instance, Sanchez-Parcerisa et al.~\cite{sanchez-parcerisa_range_2016} proposed several approaches for selecting mono- and bi-energetic arcs based on the water-equivalent range relative to the distal and proximal limits of the target. However, their method exhibits limitations when used for complex target geometries. The SPArc method \cite{ding_spot-scanning_2016} uses pre-defined beam angles, iteratively increases the sampling frequency of beam angles, and reallocates ELs to newly formed angles. Its value was demonstrated for several treatment sites \cite{li_improve_2018,liu_is_2021,liu_lung_2021,chang_feasibility_2020,ding_improving_2019,liu_improve_2020}, though computation resources suffer from the systematic re-computation of beamlets at each splitting and filtering steps. Based on geometrical target coverage, RaySearch Laboratories (Stockholm, Sweden) published a method of early energy layer selection and spot assignment (ELSA) that is implemented in RayStation \cite{engwall_fast_2022}. To our knowledge, no organ-at-risk (OAR) information is taken into account in this pre-optimization step unless the user specifies OARs through which spots are forbidden to enter. Recently, Cao et al.~presented the so-called IMPAT planning technique \cite{cao_intensity_2023} that computes prior to the geometry-based EL pre-selection, the major scanning spot contributions to the target volume. Unlike these approaches, which all rely on pre-defined gantry angles, Qian et al. introduced SPArc-particle swarm, which optimizes WET-based sector selection through particle swarm optimization to minimize energy layer switching and improve delivery efficiency \cite{qian_novel_2024}. However, none of these methods explicitly model the beam delivery time in the optimization process.


This work aims to present a new method of pre-selecting EL before robust spot weight optimization, in an efficient though flexible way, reducing the high computational load associated with full DOF optimization. The algorithm employs a bi-objective function that evaluates two competing objectives, a direct approximation of beam delivery time (BDT) and the geometric target coverage considering the location of OARs. The method also benefits from a parallelization paradigm, allowing for either faster execution or running multiple instances of the problem with different parameters. The generated solutions populate a Pareto front, providing the planner with a spectrum of trade-off solutions to choose from, offering flexibility that can be tailored both to planner preferences and specific patient needs. All the results are benchmarked against IMPT and ELSA plans. The EL pre-selection code is implemented in OpenTPS \cite{wuyckens_opentps_2023}, an open-source research treatment planning system (TPS).

\section{Materials and Method}

In this paper, PAT planning is divided into two distinct steps, following an approach that is similar to \cite{engwall_fast_2022,cao_intensity_2023}. The first step, which serves as the primary focus of this study, involves the pre-selection of EL for the arc plan. The second step entails the conventional robust spot weight optimization process, which is used to achieve dose-based objectives. In this section, we describe the EL pre-selection optimization problem and its solvers. The selected patients are then presented as well as the comparison strategy established for the study.

\subsection{Energy layer pre-selection problem}

The primary objective of the pre-selection process is to determine the optimal sequence of gantry angles and associated water-equivalent thickness (WET) values, eventually converted into energy values, that simultaneously maximizes the geometrical target coverage while minimizing the BDT. 

\begin{figure}[h]
    \centering
    \includegraphics[width=\textwidth]{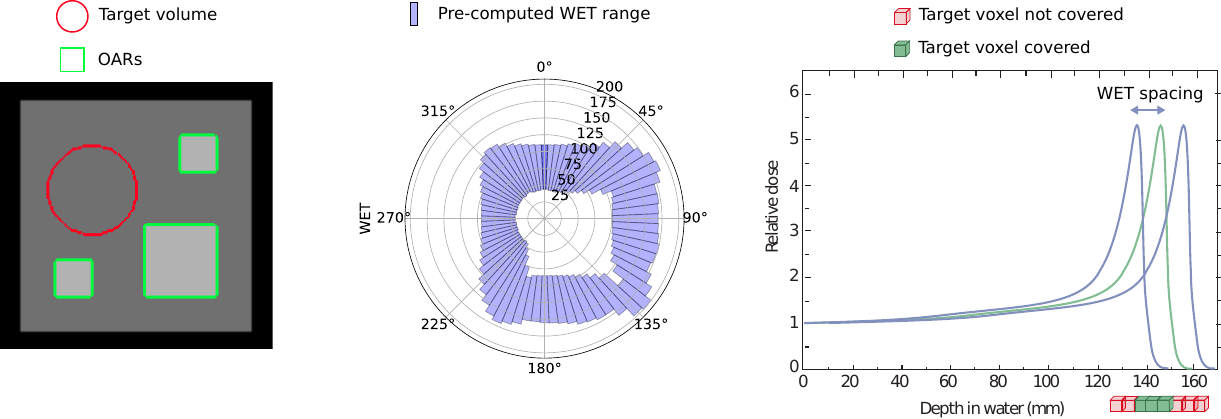}
    \caption{Left: Phantom CT example. Middle: Pre-computation of WET map with OAR considerations for a pre-defined sequence of gantry angles and WET sequence candidate found. Right: For each angle-layer combination selected, all target voxels that have a $\text{WET}_{\text{layer}} - \text{WET spacing} < \text{WET} < \text{WET}_{\text{layer}}$ are considered as covered voxels.}
    \label{fig:targetCov}
\end{figure}

During the pre-selection step, we initiate by pre-computing the WET maps $\mathcal{M}_{\text{wet}}$ through a ray-tracing procedure from the target voxels in the planning CT for a sequence of gantry angles within the user-specified arc range. Figure \ref{fig:targetCov} illustrates the WET range computed for a full arc rotation in a phantom case.

The arc can be divided into sectors, with each sector representing an arc irradiation window, during which the energy can only decrease. At each sector transition, energy can increase and carry out the so-called energy switch-up (SU). In the problem statement, the decision variables are determined by the angular positions of each sector (i.e., the starting angle $a_i$ and the angular span of each delivery window $w_i$). This approach significantly reduces the number of variables in the problem compared to selecting a decision variable for each possible gantry angle in the arc. For $M$ sectors, we consider the input solution $\bm{x} = \{ a_i, w_i\}_{i=1}^{M}$, where $a_i \leq a_{i+1}$ for $i \in \mathbb{N}$, $a_i \in [\text{minArcRange}, \text{maxArcRange}]$ and $w_i \in [\text{minWindowSize}, \text{maxWindowSize}]$ for a clockwise rotation.

The function to be optimized for pre-selection is then formulated as a multi-objective problem, divided into two parts, as shown in \eqref{eq:covObj}-\eqref{eq:timeObj}. The first part, $f_{\text{coverage}}$, is a weighted sum of several terms computed from the geometrical target coverage $\mathcal{C}$ over the $N$ voxels in the 3D CT image. To compute $\mathcal{C}$, we first need to know the sequence of angles in the arc. It can be easily inferred from utilizing the user-defined gantry spacing and the input solution $\bm{x}$. Next, we obtain the associated sequence of WET by linearly arranging WET values from its maximum possible value at the starting angle to the minimum possible WET value at the last angle in the given sector with WET values obtained by interpolation from the look-up table $\mathcal{M}_{\text{wet}}$. 

The target coverage $\mathcal{C}$ for each voxel is computed by accumulating the number of times it is covered across all gantry angles, based on whether it falls within the angle-specific WET band, which represents the Bragg Peak width and is defined by user-defined WET spacing, have their coverage incremented accordingly, as illustrated in the right plot of Fig. \ref{fig:targetCov}). If the voxel is not hidden by any OAR, its coverage is increased by 1.0. However, if it is partially obstructed by an OAR, a value between 0.0 and 1.0 is added, where the OAR-crossing penalty depends on the path length through the OAR. Note that target voxels hidden by an OAR are not included in the WET range calculation for each irradiation angle, and thus do not contribute to the target coverage.


From $\mathcal{C}$, we derive metrics such as the 10$^{th}$ percentile $\text{C}_{\text{low}}$, the minimal target coverage over the target voxels $\text{C}_{\text{min}}$. The exponential terms were chosen as objective functions to obtain positive values:

\begin{equation}
    f_{\text{coverage}}(\bm{x}) = w_{\text{low}} \cdot \exp(-\text{C}_{\text{low}}(\bm{x})) + w_{\text{min}} \cdot \exp(-\text{C}_{\text{min}}(\bm{x})) \enspace ,\label{eq:covObj}
\end{equation}

where
\begin{itemize}
    \item $\text{C}_{\text{low}}(\bm{x}) = \text{percentile}(\mathcal{C},10)$,
    \item $\text{C}_{\text{min}}(\bm{x}) = \text{min}(\mathcal{C})$,
\end{itemize}


The second objective, the BDT $T(\bm{x})$ (Eq. \ref{eq:timeObj}), is approximated by summing the irradiation time of each EL, the downward energy switching time between layers of a sector and the maximum between upward energy switching time and rotation time of the gantry between sectors. For our study, we considered a maximum gantry speed of 5 degrees per second, a fixed irradiation time of 1 second per layer, an upward energy layer switching time (ELST) of 5.5 seconds, and a downward ELST of 0.6 seconds. We included additional terms in the cost function to ensure that the minimal and maximal arc spans allowed are within specified bounds. They are formulated as semi-quadratic terms.

\begin{equation}
        f_{\text{time}}(\bm{x}) = w_{\text{T}}\cdot T(\bm{x}) + \max(0,\text{S}(\bm{x}) - \text{maxSpan})^2 
    + \min(0,\text{S}(\bm{x})- \text{minSpan})^2 \enspace ,
    \label{eq:timeObj}
\end{equation}

where
\begin{itemize}
    \item $T(\bm{x})$: BDT proxy,
    \item $S(\bm{x})$: Total arc spanning.
\end{itemize}

The weights that are integrated into the cost function were empirically determined for the first case and re-used for the others. For this study, specific values were used, which are as follows: $w_\text{low} = 10^3$, $w_\text{min} = 10^4$, and $w_\text{T} = 10^{1}$. To maintain a balanced optimization process, we further increase the importance of the total coverage objective by multiplying it by a factor of 100. This adjustment ensures that the time-related objective function does not dominate the optimization, preventing the generation of unrealistic solutions with extremely short BDT but inadequate target coverage.

The cost function formulation as described in \eqref{eq:covObj}-\eqref{eq:timeObj} allows for fast evaluation and motivates this study. Compared to a dose-based EL selection approach, not only does it reduce the number of optimization variables (spot weights $\rightarrow$ irradiation windows), it also significantly lowers the number of constraints in the objective formulation by replacing the heavy dose computation with fast geometrical target coverage and delivery time evaluation. To ensure clarity throughout the paper when discussing the EL pre-selection method developed in this paper, we will use the term `Flexi-Arc'.

\subsection{Optimization algorithms}

The two competing objectives could be combined into a single objective function using a weighting factor. The main difficulty is that the objective function is not convex at all, so many local minima can exist. Furthermore, the space of solutions is large and multi-dimensional, even if it has bounds.
To address these challenges, we selected global multi-objective methods based on evolutionary computation, utilizing the PyGMO (Python Parallel Global Multiobjective Optimizer) \cite{biscani_parallel_2020}, a Python scientific library for massively parallel optimization. 

For our study, we utilized the Generalized Island Model \cite{fernandez_de_vega_generalized_2012}, a key tool in the PyGMO library, for solving complex optimization problems. This model enables parallel optimization by partitioning a population (i.e., candidate solutions to the given problem) into multiple `islands', each operating independently with distinct optimization algorithms and/or parameters. Periodic migration of population individuals, between islands, fosters diversity and helps explore the search space, ultimately leading to improved convergence properties. Our investigation focused on several multi-objective algorithms available in PyGMO: NGSA2 \cite{deb_fast_2002}, MOEAD \cite{qingfu_zhang_moead_2007}, NSPSO \cite{hui_li_multiobjective_2009}, and MACO \cite{acciarini_mhaco_2020}. The islands evolve asynchronously (i.e., running as `background' tasks), each within a separate process, thread, or machine. 

\begin{figure}[h!]
    \centering
    \includegraphics[width=\textwidth]{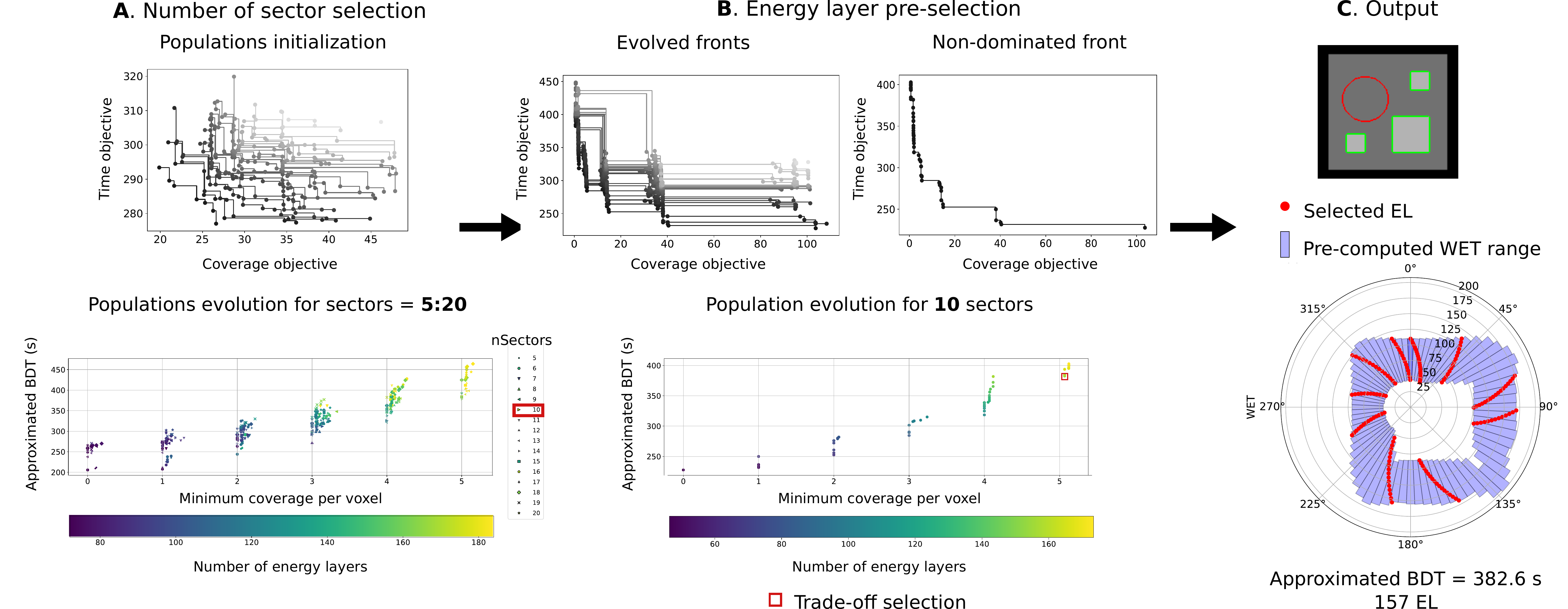}
    \caption{Energy layer pre-selection workflow implemented using PyGMO \cite{biscani_parallel_2020}. In Step A, we initialize a population for each number of sectors, allowing them to evolve towards convergence on their respective Pareto fronts. We select the 10-sector Pareto front as a suitable trade-off. In Step B, we continue the evolution of the 10-sector population. After filtering and retaining the non-dominated front, we choose a solution out of it. Finally, in Step C, we present the solution, i.e., the selected ELs to be used in the dose optimization process. }
    \label{fig:workflow}
\end{figure}

Figure \ref{fig:workflow} depicts the workflow of the algorithm. It is executed in two main phases. In the first phase (A), we determine the number of sectors required to achieve proper geometrical target coverage. For that purpose, multiple processors run instances of the problem independently with varying numbers of sectors, e.g., ranging from 5 to 20 sectors for a full-arc rotation. Each island converged towards a non-dominated front corresponding to a specific number of sectors. By comparing the obtained Pareto fronts, we assessed various trade-offs, enabling users to select the optimal number of sectors based on time and coverage efficiency.

In the second phase (B), we aim to refine the previously obtained Pareto front. For this purpose, we run the same problem instance (with the previously selected number of sectors) in parallel on different processors using a fully-connected topology. In this setup, islands are all interconnected and can exchange individuals between each other. After evolving their populations, the islands merge the solutions into several non-dominated fronts. The fast non-dominated sorting algorithm available in PyGMO is then employed to retain only the solutions present on the highest ranked non-dominated front. Finally, we manually filter solutions with time objectives twice as the minimum time objective among the solutions, thus yielding a realistic set of optimal solutions. The final EL pre-selection is ultimately chosen based on the non-dominated front solutions displaying the BDT versus target coverage metric (e.g., the minimal target or $10^{th}$ percentile coverage). The final output (C) is a list of the gantry angles and for each of the associated ELs that were found. 

The algorithm is integrated and openly available in OpenTPS \cite{wuyckens_opentps_2023}, an open-source python-based TPS for research in proton therapy. Simulations were performed on an Intel(R) Xeon(R) Gold 6248 CPU @ 2.50GHz with a total of 80 CPU cores and 526 GB RAM. The computation time is highly dependent on the algorithm parameters.  With the chosen parameters, the optimization process consistently took around 5 minutes on average to generate the Pareto front for EL pre-selection.

The resulting list of ELs and corresponding gantry angles could now be imported in any TPS capable of selecting appropriate spot positions for a user-defined set of ELs. Once the spot map corresponding to each of these layers is computed, one can proceed with the conventional robust spot weight optimization.

\subsection{Patient data}
We applied the previously described method to three different treatment sites. The first case involves a young adult patient with craniopharyngioma. The anonymized patient was retrieved from a database of the Cliniques Universitaires Saint-Luc (Brussels, Belgium), for which ethical approval was granted. The prescription for this craniopharyngioma is 54 Gy in 30 fractions. The second case consists of a patient with locally advanced esophageal cancer. The use of this patient data for the study was approved by the Institutional Ethical Review Board of the UZLeuven (S59667, Leuven, Belgium). The patient was planned with a prescribed dose of 50.4 Gy delivered in 28 fractions. Finally, the third case is a patient with non-resectable lung cancer located in the right middle lobe. The prescribed dose was 60 Gy in 30 fractions. This patient belongs to a database of the Cliniques Universitaires Saint-Luc (Brussels, Belgium), for which ethical approval was also granted.

\subsection{Treatment planning}

Treatment planning optimization was done in a research version of RayStation 2023B (Raysearch Laboratories, AB). All plans were benchmarked against ELSA and IMPT plans. The proton therapy system used for planning was a Proteus Plus (IBA s.a., Belgium).

For the craniopharyngioma case, an IMPT plan was generated with 3 beams including a right–anterior–superior beam (gantry [G] $=70^{\circ}$, couch [C] $=220^{\circ}$), a left–anterior–superior beam ([G] $=70^{\circ}$, [C] $=320^{\circ}$), and an oblique vertex beam ([G]$=315^{\circ}$, [C] $=90^{\circ}$). Several proton arc plans were designed with a full 360$^{\circ}$ arc. The ELSA algorithm used in this study slightly differs from the original publication \cite{engwall_fast_2022} in the sense that it pre-selects ELs with a more predefined sector length, which explains the consistency in the number of sectors and associated energy SUs in the plans. Several gantry angle spacing values were tested for the ELSA plans: we designed plans with 2$^{\circ}$, 3$^{\circ}$ and 4$^{\circ}$ as gantry angle spacing, corresponding to 180, 120 and 90 ELs. This matched with using 8, 5 and 5 sectors, resulting in 7, 4 and 4 SUs, respectively. The Flexi-Arc plans were generated with a 1-degree angle spacing to give more freedom to place sectors and their EL in the full arc rotation. The number of sectors and EL for Flexi-Arc is defined based on the trade-off solution picked on the Pareto fronts produced and is studied in Section \ref{seq:results}. WET maps were pre-computed for each angle as well as the corresponding WET range taking into account two selected OARs: the optic nerves and the hippocampi. The WET range was therefore slightly shrunk for angles that cross these OARs to reach the target voxels. 
Regarding the esophageal cancer case, two posterior beams [G] = 150°,180°, [C] = 0°,0°) on the average CT scan used as reference for planning were set for IMPT planning, following the PROTECT guidelines \cite{HOFFMANN202232}. A single ELSA plan was made with a full arc design using 2° angle spacing and therefore 180 ELs. The Flexi-Arc plan was parameterized to be able to use the full arc range with a 1° angle spacing. The heart was designated as the OAR penalty in the algorithm. For the lung cancer case, three beams were used in the IMPT plan at [G] = 190°, 240° and 290° and [C] = 0°. Both PAT plans (ELSA and Flexi-Arc) were designed using a partial arc, to spare the contralateral healthy lung. In this case, the heart was again chosen as the penalty OAR in Flexi-Arc energy layer pre-selection. A summary of the different beam configurations used for the different plans and modalities is provided in the Supplementary Material (SM) 1. The patient anatomy and pre-computed WET ranges for the three selected patient are displayed in SM 3.

After the EL pre-selection (either ELSA or Flexi-Arc), Monte Carlo dose calculation and robust worst-case optimization \cite{fredriksson_minimax_2011} were carried out for each patient. The same settings were used to generate both the IMPT and PAT plans.  In the craniopharyngioma case, 21 scenarios were considered (3\% proton range error and 3 mm set-up isotropic error) for the CTV and brain stem. Regarding the esophageal tumor, an iCTV was defined by delineating the CTV in all breathing phases, then mapping the contours to the average CT before computing their union. 45 scenarios (2.6\% range error and 7 mm setup error) were considered in the worst-case (minimax) robust optimization, applying robust objectives on the iCTV and on the spinal canal. For the lung tumor, the robust optimization included a total of 84 scenarios: 7 setup error scenarios (5 mm in 6 different directions, plus nominal case) times 3 range scenarios (range error, plus nominal case) times 4 phases (MidP, EndExh, EndInh, MidV). More details on the patient-specific IMPT plans can be found in a prior study \cite{chocan_dosimetric_2024}.

To summarize, the Flexi-Arc planning workflow consists of (i) looking for an optimal EL selection using a PyGMO algorithm, (ii) optimizing the plan within RayStation, and (iii) comparing the clinical goals to those of ELSA and IMPT plans.


\subsection{Plan evaluation}

Clinical goals are provided in SM 3 for each considered treatment site. Several other metrics of interest were also computed such as the homogeneity index (HI) as defined in ICRU 83, the conformity index (CI) recommended in
the RTOG radiosurgery guideline \cite{shaw_radiation_1993}, and the body integral dose (ID). Additionally, a robust evaluation of each plan was performed. For the esophageal cancer and lung cancer cases, we could extend the comparison study by evaluating heart and lungs complications after definitive chemoradiotherapy. For this purpose, we chose the Lyman-Kutcher-Birman (LKB) normal tissue complication probability (NTCP) model \cite{KUTCHER19891623}. We focused on toxicities inflicted to the lungs (pneumonitis) and heart (pericardial
effusion). More details on the model and its chosen parameters can be found in prior studies \cite{chocan_dosimetric_2024,Wera2024}.

Finally, while we use a very rough proxy of the BDT taken by the ELs coming out of the pre-selection stage, we would like to use a better estimation of the BDT for the final plan comparison. Therefore, we computed the dynamic BDT with the ATOM\footnote{\url{https://github.com/raysearchlabs/ArcTrajectoryOptimizationMethod}} (Arc Trajectory Optimization Method) algorithm, an open-source tool \cite{wase_optimizing_2024} published by RaySearch Laboratories. ATOM is designed to determine a fast plan delivery while adhering to the mechanical constraints specified by the user. The machine parameters utilized in this study are reported in Table \ref{tab:machineParam}. Relying on ATOM was driven by its utility, given that proton arc has not been commissioned for clinical use yet.

\begin{table}[h!]
\centering
{\fontsize{10}{12}\selectfont
\begin{tabular}{lc} 
\toprule
\textbf{Maximum gantry velocity} & 5°/s \\
\textbf{Maximum gantry acceleration} & 0.5 °/s$^2$  \\
\textbf{Maximum gantry jerk} & 0.5 °/s$^3$ \\
\textbf{Downward energy switching time} & 0.6 s      \\ 
\textbf{Upward energy switching time} & 5.5 s        \\
\textbf{Spot switching time} & 2 ms       \\
\textbf{Spot delivery time (per MU)} & 5 ms       \\ \bottomrule
\end{tabular}%
}
\caption{Machine constraints parameters used for BDT estimation by ATOM \cite{wase_optimizing_2024}.}
\label{tab:machineParam}
\end{table}

For clarity, when using ATOM to compute the BDT, we will refer to the \textit{ideal} BDT while we will refer to the BDT \textit{proxy} for its quick and rough estimation during the pre-selection stage.

\section{Results}
The results of EL pre-selection are detailed for the craniopharyngioma case to give readers a deeper understanding of how the Flexi-Arc plans were designed. For the other two treatment sites, we focus on dosimetry and toxicity comparisons to ELSA plans, without delving into the specifics of EL pre-selection. For conciseness, the comparison to IMPT plans is provided in SM 5 and 6.

\subsection{Craniopharyngioma}

\begin{figure}[h!]
    \centering
    \includegraphics[width=\textwidth]{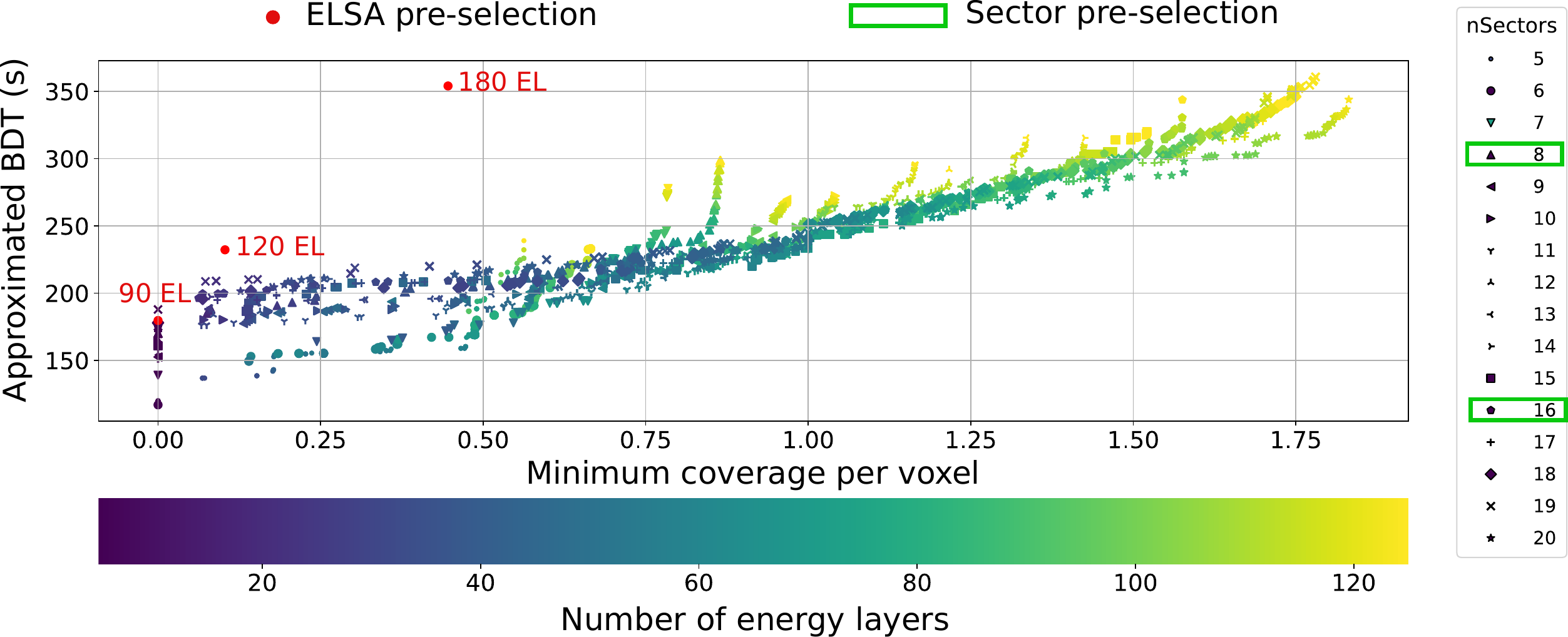}
    \caption{Non-dominated fronts for several numbers of sectors. Minimum coverage and BDT increases with the number of sectors. ELSA solution set, which included 90, 120 and 180 ELs, was also evaluated in relation to these two objectives (though not optimized using the exact same definition).}
    \label{fig:xSectors}
\end{figure}

Figure \ref{fig:xSectors} shows the Pareto fronts, which depict the trade-offs between the approximated BDT and the minimum target voxel coverage. We solved 21 parallel instances of the problem, varying the number of sectors from 5 to 20, hence the 21 Pareto fronts. The observed trend indicates that as we increase the number of sectors, the Pareto front shifts upward, meaning improved target coverage but longer BDT. Interestingly, it is the total number of EL that has a more substantial impact on BDT than the number of sectors. Moreover, the use of more sectors allows for a greater number of energy layers to be incorporated (as angle spacing is fixed), thereby enhancing target coverage. We also assessed the ELSA EL pre-selections regarding these two metrics employed for Flexi-Arc EL pre-selection (indicated by the red dots in Fig.~\ref{fig:xSectors}). It is noteworthy that while ELSA pre-selections may not appear Pareto-optimal on this plot, the comparison is not entirely fair because ELSA has been optimized using a different definition of these criteria. Consequently, the `worse' value observed for ELSA were somewhat expected. We selected two Pareto fronts to continue the study: the 8-sector front, to allow a direct comparison with the 8-sector ELSA pre-selection, and the 16-sector front, which is shifted further to the right on the minimum coverage axis and includes data points with low BDT proxys (not achievable with the ELSA method).


\label{seq:results}

\begin{figure}[h!]
    \centering
    \includegraphics[width=\textwidth]{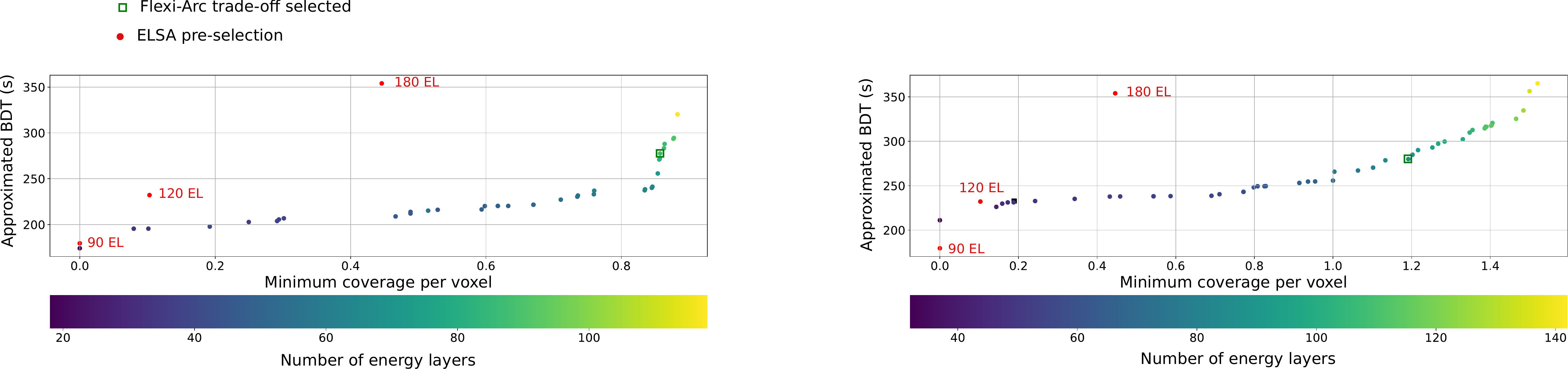}
    \caption{Non-dominated front refined for 8 sectors (left) and 16 sectors (right). Minimum coverage per voxel and BDT increases with the total number of EL. Final trade-off EL pre-selected  in a green square. }
    \label{fig:8-16Sectors}
\end{figure}

Next, we ran and solved separate instances of the 8-sector and 16-sector problems. After filtering out non-realistic solutions and retaining only the highest-ranked non-dominated solutions, we get the Pareto fronts, as shown in Fig.~\ref{fig:8-16Sectors}. As anticipated, the 16-sector solution achieves a higher minimum target coverage compared to the 8-sector solution. The 8-sector front completely dominates the ELSA pre-selections. The Pareto fronts in these plots clearly show that the BDT increases as the total number of ELs increases. We carefully selected one data point from each of these Pareto fronts, taking into account not only achieving a reasonable BDT and a high minimum coverage but also conducting individual visual inspections of the EL layout on a polar plot to avoid configurations with unique EL within sectors, for instance. The retained EL pre-selections include 87 and 85 ELs for the 8- and 16-sector solutions, respectively.

\begin{figure}[h!]
    \centering
    \includegraphics[width=\textwidth]{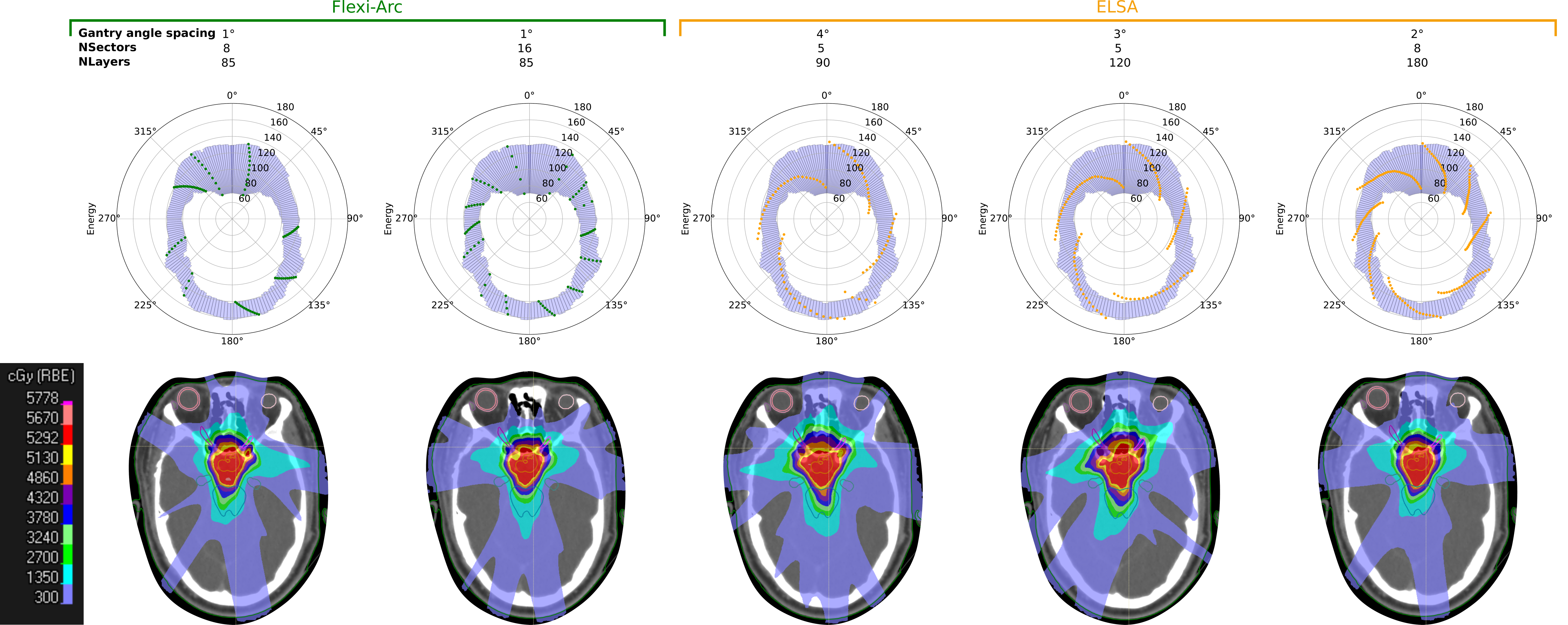}
    \caption{Energy layer pre-selection for each trade-off selected using the Flexi-Arc method and for each ELSA plan. Final dose distributions are displayed below after spot weight optimization in the RayStation TPS.}
    \label{fig:polarBrain}
\end{figure}

Figure \ref{fig:polarBrain} displays information on two rows. In the first one, we show the EL pre-selection results that the two different methods yielded: the Flexi-Arc method for both 8- and 16-sector configurations and the ELSA method for various gantry angle spacing values. The Flexi-Arc pre-selection naturally spaces out the sectors, accommodating dead angles, useful when potentially crossing the selected sensitive structures (optic nerves and hippocampi) and minimize gantry braking to accommodate the SUs. In contrast, ELSA exhibits a more regular pattern with no consideration for OAR avoidance and no gaps between sectors. The occurrence of some ELs falling outside the WET range calculated for the ELSA pre-selection could be explained by differences in the WET ray-tracing implementation and by the fact that ELSA accounts for range errors already in the EL selection while Flexi-Arc accounts for the position of selected OARs.

In the second row of Fig.~\ref{fig:polarBrain}, we present the dose distributions obtained after optimizing the spot weights for each of these EL pre-selection methods within RayStation. The dose optimization was performed individually for each plan to maximize their quality. It seems from these dose distributions that the best target conformity was achieved with the 16-sector Flexi-Arc and the 2-degree spacing ELSA plans. The 16-sector Flexi-Arc plan effectively avoids anterior irradiation to attempt to spare the optic nerves and employs oblique angles to protect the hippocampi as intended.

In Table \ref{tab:ctvCov}, we report plan quality metrics. The dose was normalized so that D98 corresponds to the CTV prescription, meeting the clinical goal outlined in Table \ref{tab:clinicGoals} from SM 3. As anticipated, the highest CI is achieved with the 2° ELSA plan and the 16-sector Flexi-Arc plan. The CI deteriorates for the 3° and 4° ELSA plans, resulting in a higher integral dose to the body. The best HI (lowest) was obtained with the ELSA 2° plan, likely due to the increased number of ELs. Additionally, the number of spots is significantly higher for the ELSA plans. When comparing the ideal BDT among the PAT plans, the Flexi-Arc plans achieve the shortest time, aligning with the reduced number of ELs needed for target coverage and the gap between sectors. For instance, the 16-sector Flexi-Arc plan achieves a reduction of 16\%, 18\% and 30\% compared to the 4°, 3°, and 2° ELSA plans, respectively.

In terms of a comparison with the IMPT plan, it is noteworthy that the IMPT plan already exhibited very high quality, with an even lower integral dose as compared to the PAT plans.

\begin{table}[h!]
\centering
{\fontsize{10}{12}\selectfont
\begin{tabular}{@{}lcccccc@{}}
\toprule
\textbf{} & \multicolumn{2}{c}{\textbf{Flexi-Arc}} & \multicolumn{3}{c}{\textbf{ELSA}} & \textbf{IMPT} \\ 
\cmidrule(l){2-3} \cmidrule(l){4-6} 
 \textbf{Gantry angle spacing ($^{\circ}$)} & 1 & 1 & 4 & 3 & 2 & \\
 \textbf{NSectors}         & 8   & 16 & 5 & 5 & 8  & \\ 
\midrule 
D98(CTV)-nom (Gy [RBE])      & 54.0 & 54.0 & 54.0 & 54.0 & 54.0 & 54.0\\
D98(CTV)-wc (Gy [RBE])       &  53.2& 53.2&   53.4 & 53.5 & 53.5 & \textbf{53.8}\\
D2(CTV)-nom (Gy [RBE])   & 56.3 & \textbf{56.0} &    56.6& 56.3 & \textbf{56.0} & 56.5\\
D2(CTV)-wc (Gy [RBE])  & 57.0 & 56.5 &  57.0 & 56.8 & \textbf{56.4} & 56.6\\
CI & 3.7 & \textbf{3.5} &   5.1 & 4.9 & \textbf{3.5} & 3.7\\
HI  & 0.041 & 0.045 &   0.047 & 0.041 & \textbf{0.037} & 0.045\\
Patient ID (Gy $\cdot$ L)    & 13.2 & 13.2 &   20.7& 20.8 & 12.5 & \textbf{10.1} \\
\midrule 
NSpots       &  2816& 2019 &  5817& 6206 & 4834 & 2620\\
NLayers  & 87 & 85 &   90 & 120 & 180 & 41\\
\textit{Ideal} BDT (s)  &  182 & \textbf{171}  & 203 & 209 &245 & / \\
\bottomrule
\end{tabular}%
}
\caption{\footnotesize Dosimetric evaluation and plan metrics for the \textbf{craniopharyngioma} case. \textit{Abbreviations:  D98(CTV) = dose received at 98\% of the CTV, nom = nominal, wc = worst-case, CI = conformity index, HI = homogeneity index, ID = body integral dose, BDT = beam delivery time obtained with ATOM \cite{wase_optimizing_2024}.}}
\label{tab:ctvCov}
\end{table}

\begin{figure}[h!]
    \centering
    \includegraphics[scale=0.5]{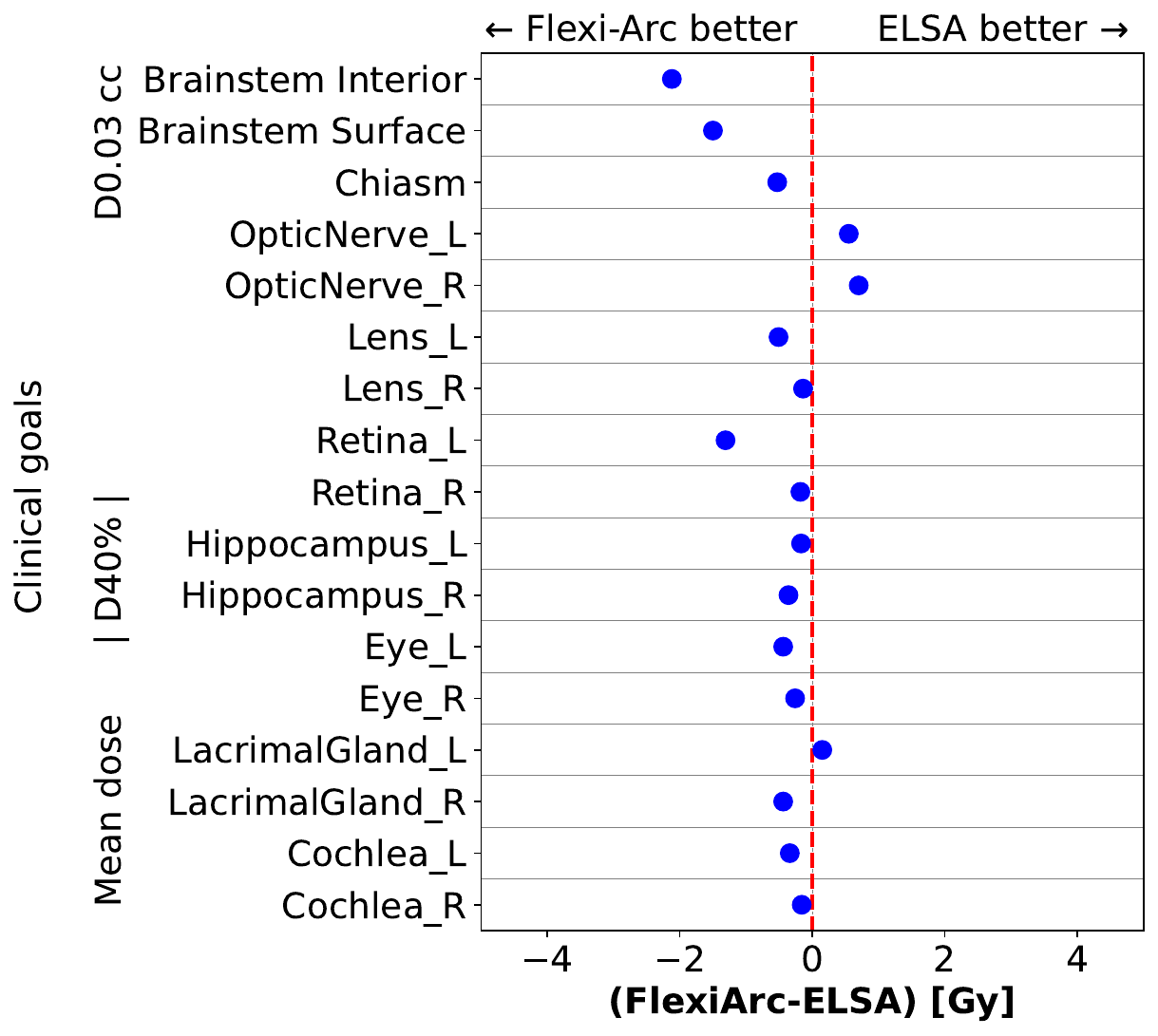}
    \caption{Differences in terms of clinical planning criteria for the OARs between the 16-sector Flexi-Arc plan and the 2° spacing ELSA plan (\textbf{brain} tumor case). Negative differences imply that the Flexi-Arc plan was better than the ELSA plan for that metric (i.e., less dose to the OARs). }
    \label{fig:OARDosiBrain}
\end{figure}

Finally, in our evaluation, we focused not only on target coverage and BDT but also on the dosimetry of OARs. Figure \ref{fig:OARDosiBrain} provides a comparison of dose metrics related to the clinical OAR goals for the Flexi-Arc 16-sectors plan and the ELSA 180 EL plan. Overall, the Flexi-Arc plans tend to better spare the OARs compared to ELSA, with the exception of the optic nerves, despite being selected as critical OARs during EL pre-selection. However, it was effective in sparing the hippocampi. Nevertheless, notice that the differences between ELSA and Flexi-Arc plans are generally quite small. A fully detailed dosimetric evaluation of the OARs in the nominal and worst-case scenario for all the plans generated is provided in SM 6. The comparison to the IMPT plan is also included.

\subsection{Esophageal tumor}

\begin{figure}[h!]
    \centering
    \includegraphics[width=0.7\linewidth]{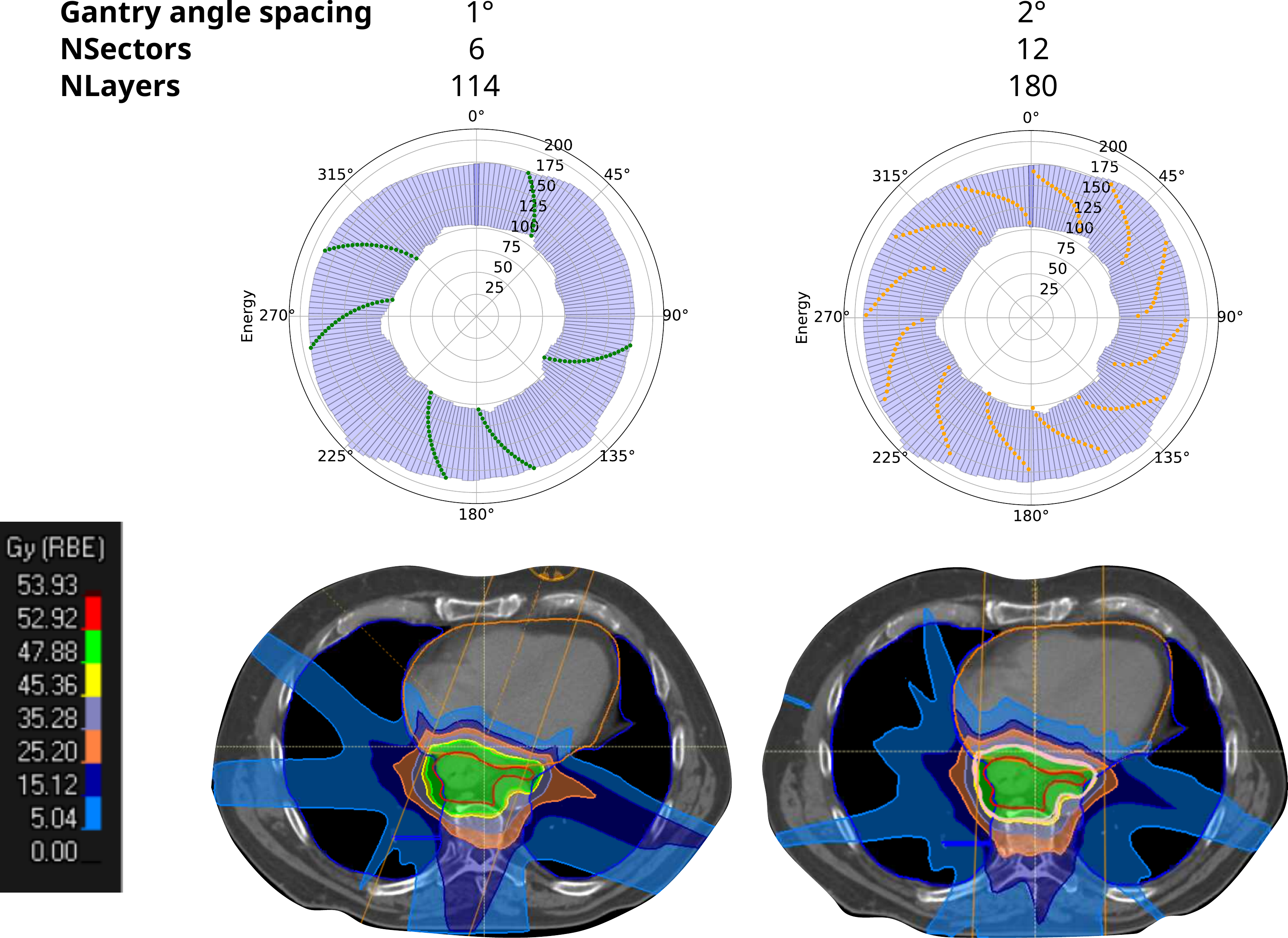}
    \caption{Energy layer pre-selection selected using the Flexi-Arc (left) and ELSA (right) methods in the \textbf{esophageal} tumor case. Final dose distributions are displayed below after spot weight optimization in the RayStation TPS.}
    \label{fig:esoDose}
\end{figure}

\begin{figure}[h!]
    \centering
    \includegraphics[width=0.7\linewidth]{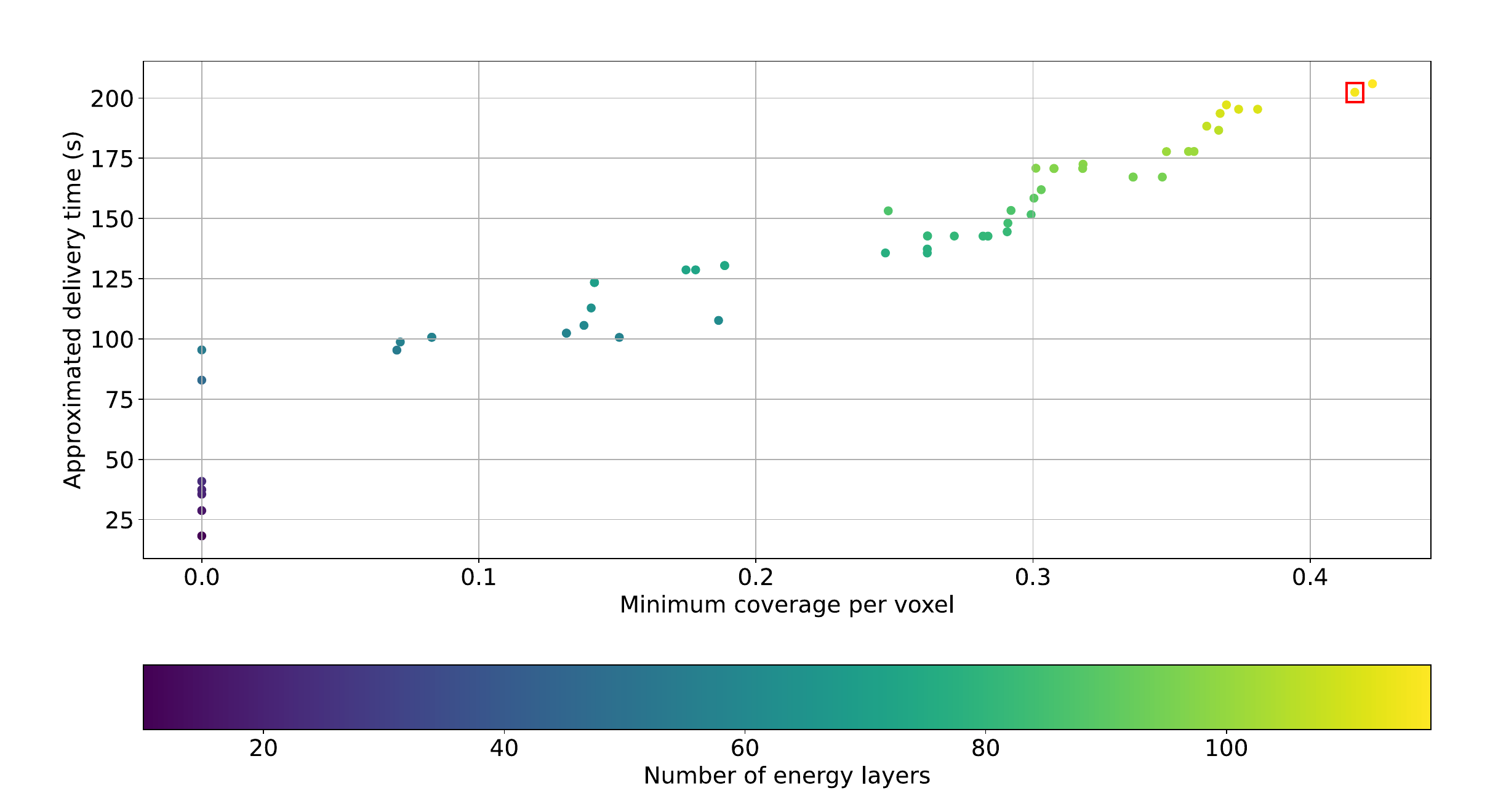}
    \caption{Pareto front for the \textbf{esophageal} tumor case (6-sectors solutions). The ``best" tradeoff selected is the red square. It corresponds to a specific energy layer pre-selection.}
    \label{fig:paretoEso}
\end{figure}

Flexi-Arc was again used to generate a non-dominated front consisting of candidate sets for EL pre-selection in the esophageal cancer case (see Fig. \ref{fig:paretoEso} in SM 4). After thorough evaluation, a single candidate set was selected, featuring 114 ELs arranged into six sectors spanning the full 360° arc, with a 1° spacing between ELs within each sector. In contrast, the ELSA plan used 12 sectors and 180 ELs. Consequently, the BDT for the Flexi-Arc plan was reduced by 41\% compared to the ELSA method (see Table \ref{tab:esoDosi} for exact numbers). Figure \ref{fig:esoDose} illustrates the ELs that get selected by both approaches and the resulting dose distributions after spot weight optimization in the RayStation TPS. The 6-sector Flexi-Arc solution resembles an IMPT plan, with `condensed' ELs within each sector and large gaps in-between sectors, while the dose distribution from the ELSA plan appears more `diffuse' due to the larger number of sectors and EL used.

\begin{figure}[h!]
\centering
\begin{subfigure}{.5\textwidth}
  \centering
  \includegraphics[width=0.9\linewidth]{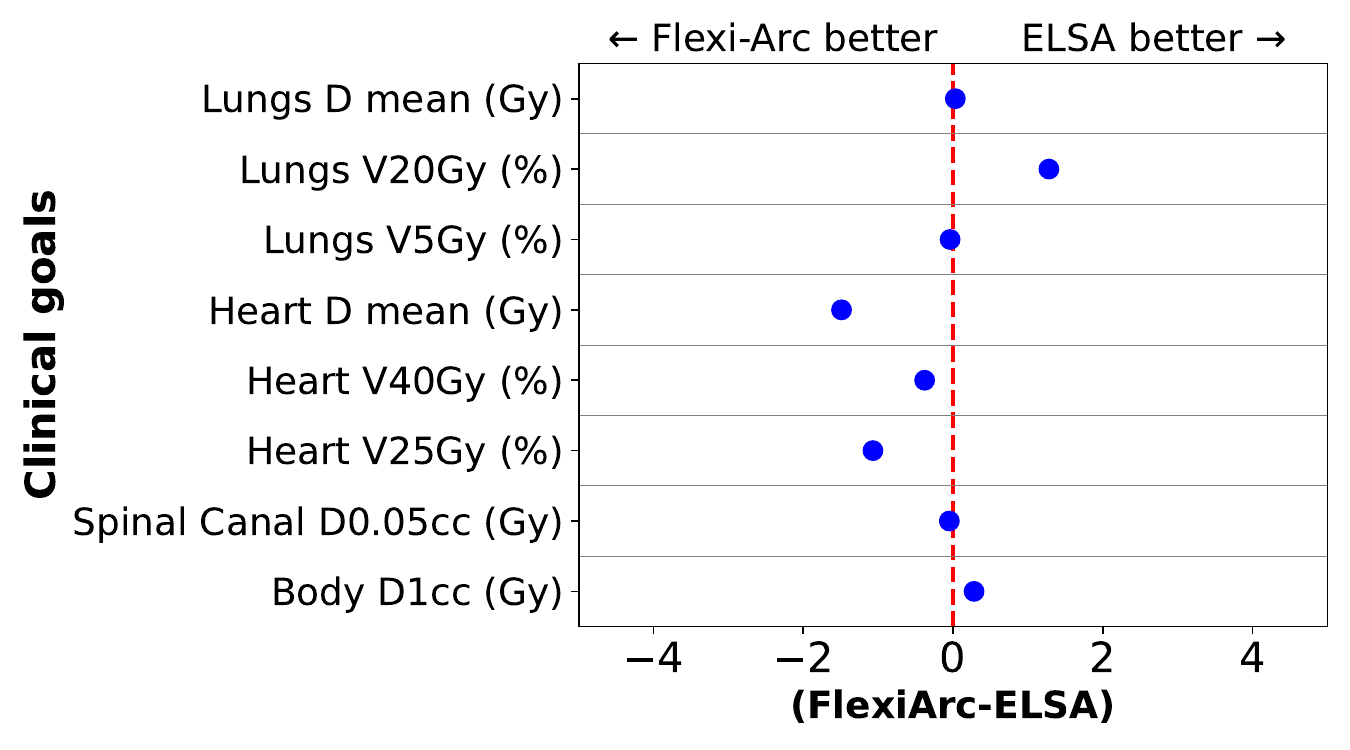}
  \caption{Differences in OARs dose}
  \label{fig:sub-oarEso}
\end{subfigure}%
\begin{subfigure}{.5\textwidth}
  \centering
  \includegraphics[width=.8\linewidth]{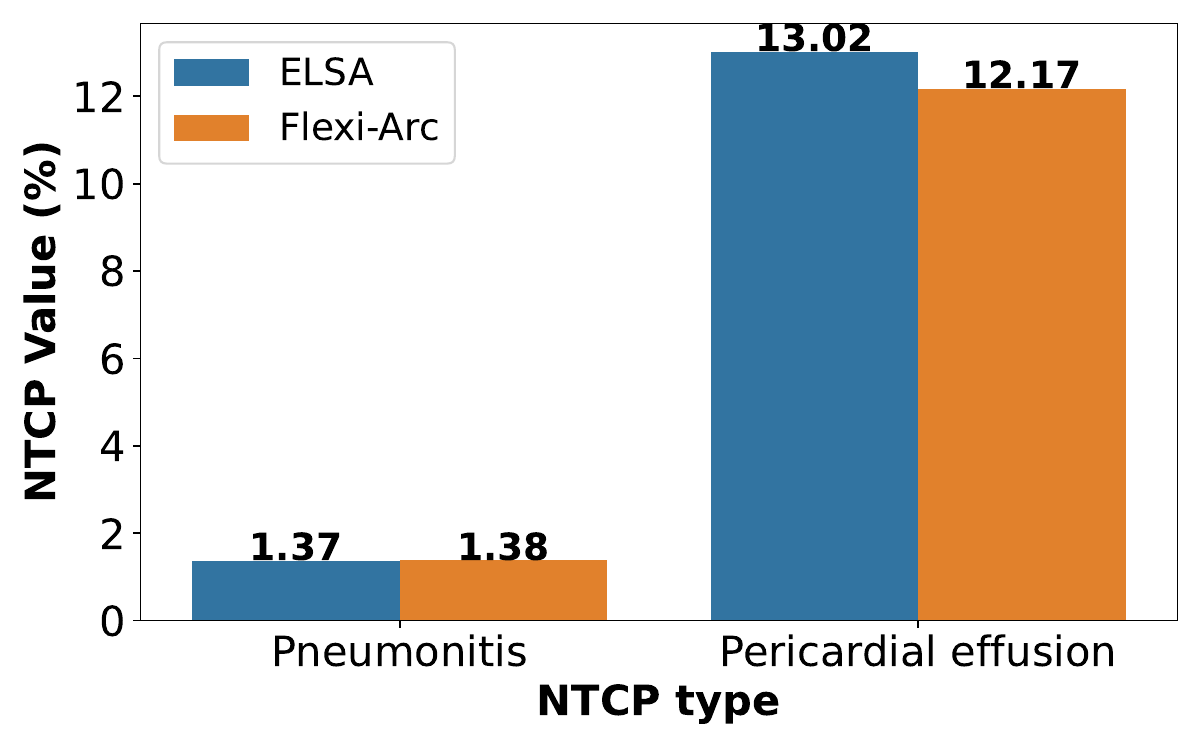}
  \caption{NTCP}
  \label{fig:sub-ntcpEso}
\end{subfigure}
\caption{Comparison of OAR doses and toxicities between the Flexi-Arc plan and the ELSA plan (\textbf{esophageal} tumor case). Negative differences imply that the Flexi-Arc plan was better than the ELSA plan for that metric (i.e., less dose to the OARs).}
\label{fig:toxicityEso}
\end{figure}

Dosimetrically, both plans provided equivalent target coverage under nominal and worst-case scenarios (Table \ref{tab:esoDosi}), as well as comparable OAR sparing (Fig.~\ref{fig:sub-oarEso}). However, the Flexi-Arc method showed a slight advantage, particularly in sparing the heart. The NTCP results (Fig.~\ref{fig:sub-ntcpEso}) showed no meaningful differences between the two planning approaches.

Compared to the IMPT plan, the Flexi-Arc plans exhibit a slight reduction in target conformity and homogeneity, although the differences are not substantial. Additionally, the integral dose may require attention with PAT, as it increases compared to conventional planning.

\begin{table}[h!]
\centering
\begin{tabular}{llll} \toprule
                             & \textbf{Flexi-Arc} & \textbf{ELSA}  & \textbf{IMPT} \\
\textbf{Gantry angle spacing (°)  }      & 1        & 2    & \multirow{2}{*}{} \\
\textbf{NSectors}                     & 6         & 12    &                   \\ \midrule
iCTV mean-nom (Gy {[}RBE{]}) & 50.4      & 50.4  & 50.4              \\
iCTV mean-wc (Gy {[}RBE{]})            & \textbf{50.4}      & 50.3  & 50.5              \\
iCTV V95\%-nom (Gy {[}RBE{]})          & 100       & 100   & 100               \\
iCTV V95\%-wc (Gy {[}RBE{]})            & \textbf{99.9}      & \textbf{99.9}   & 99.7              \\
CI                           & 2.36      & 2.15  &   \textbf{2.12}               \\
HI                           & 0.04      & \textbf{0.03}  &   0.05                \\
Patient ID ( Gy $\cdot$ L)   & 160.9     & 147.2 &   \textbf{136.9}               \\ \midrule
NSpots                       & 61113     & 95742 & 10171             \\
NLayers                      & 114       & 180   & 37                \\
\textit{Ideal} BDT (s)& \textbf{281.1}     & 480.3 & /    \\  \bottomrule             
\end{tabular}
\caption{Dosimetric evaluation and plan metrics for the \textbf{esophagus} case.}
\label{tab:esoDosi}
\end{table}

\subsection{Lung tumor}

\begin{figure}[h!]
    \centering
    \includegraphics[width=0.7\linewidth]{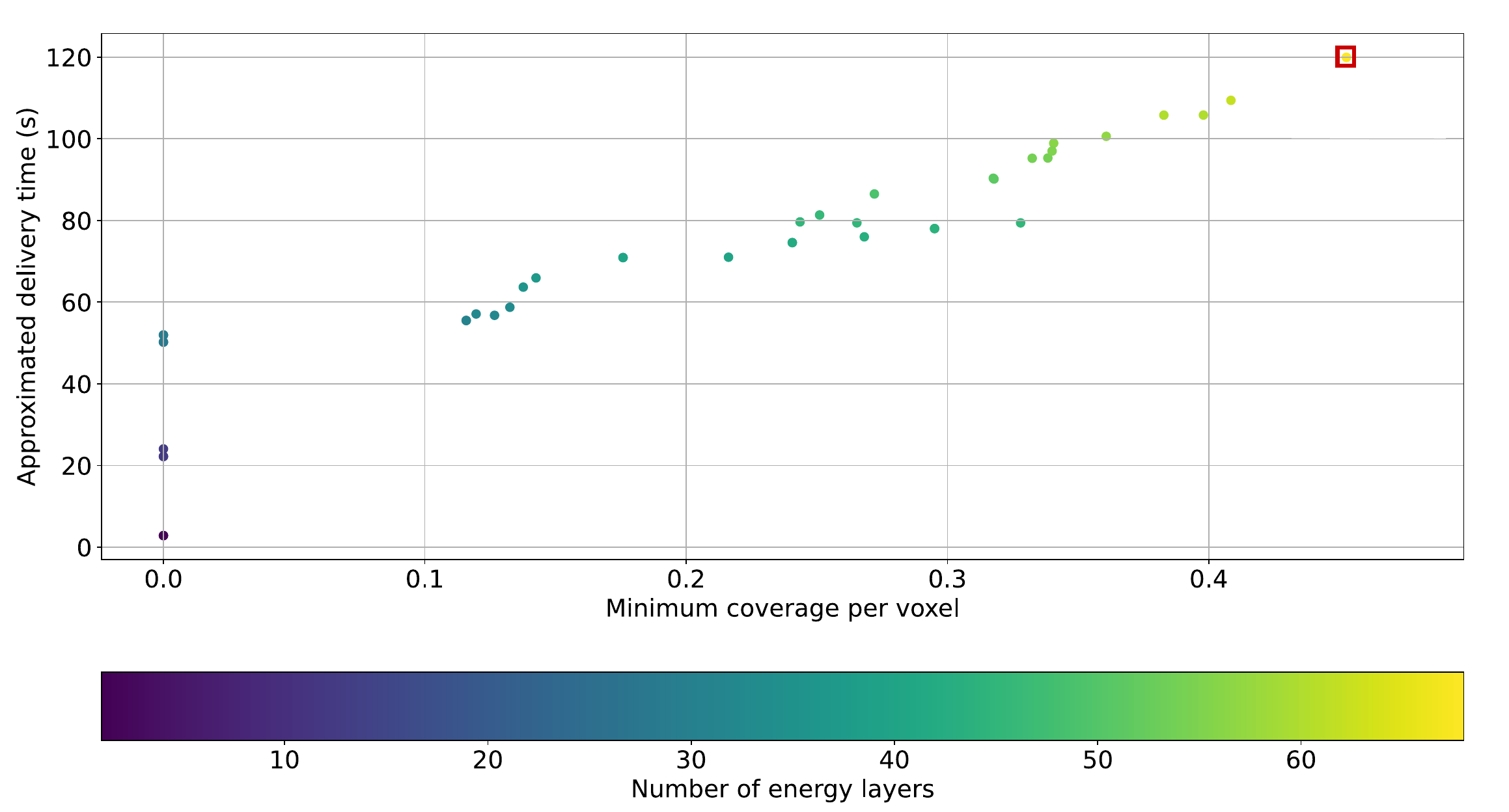}
    \caption{Pareto front for the \textbf{lung} tumor case (6-sectors solutions). The ``best" tradeoff selected is the red square. It corresponds to a specific energy layer pre-selection.}
    \label{fig:paretoLung}
\end{figure}

For the last case featuring a partial arc, a 6-sector solution with 68 EL was selected from the non-dominated front after running Flexi-Arc (see Fig. \ref{fig:paretoLung} in SM 4). This was compared to an ELSA plan designed with 5 sectors and 71 EL. As shown in Table \ref{tab:lungDosi}, Flexi-Arc resulted in a 17\% reduction in BDT compared to ELSA. In addition, this reduction in time was achieved while maintaining equivalent plan quality between both approaches, further highlighting Flexi-Arc's advantage. Strictly based on the metrics reported in Table \ref{tab:lungDosi}, Flexi-Arc demonstrates slightly better performance compared to ELSA as well as the IMPT plan with the lowest CI and integral dose.

\begin{figure}[h!]
\centering
\begin{subfigure}{.5\textwidth}
  \centering
  \includegraphics[width=0.9\linewidth]{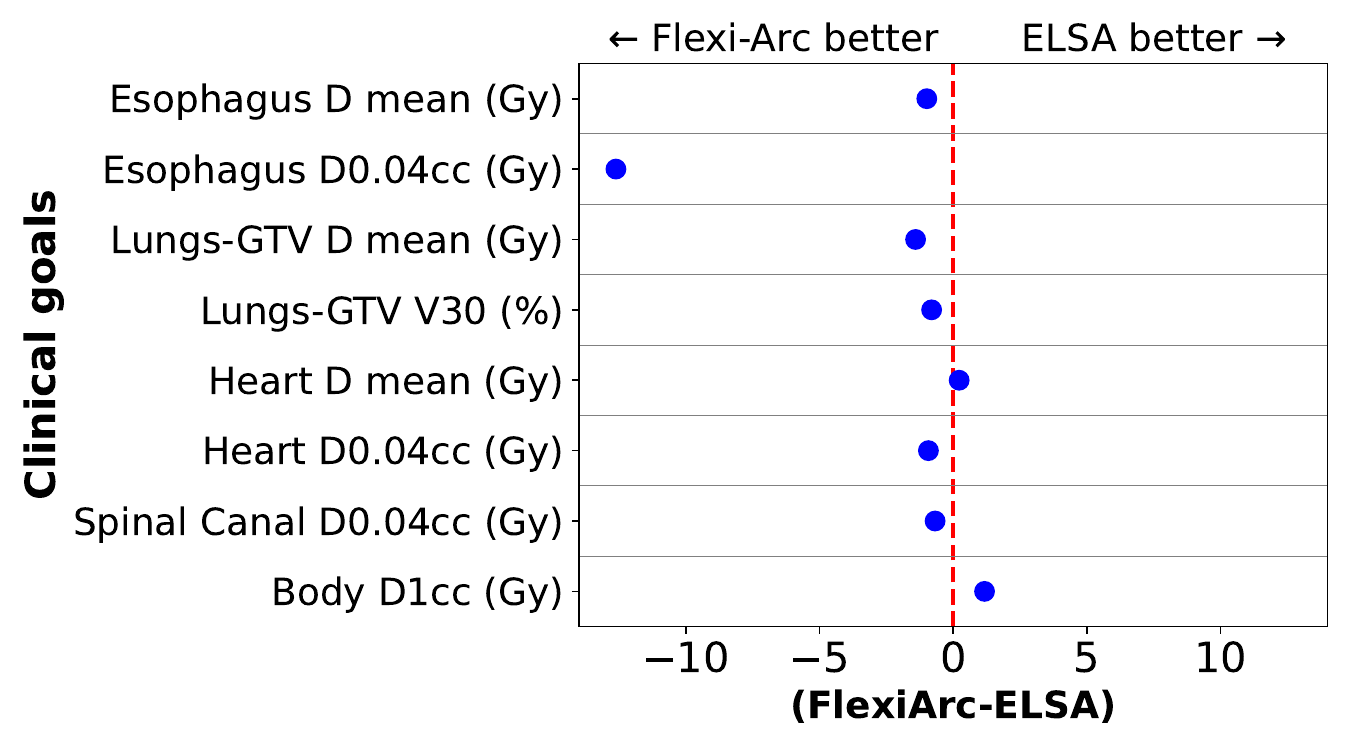}
  \caption{Differences in OARs dose}
  \label{fig:sub-oarLung}
\end{subfigure}%
\begin{subfigure}{.5\textwidth}
  \centering
  \includegraphics[width=.8\linewidth]{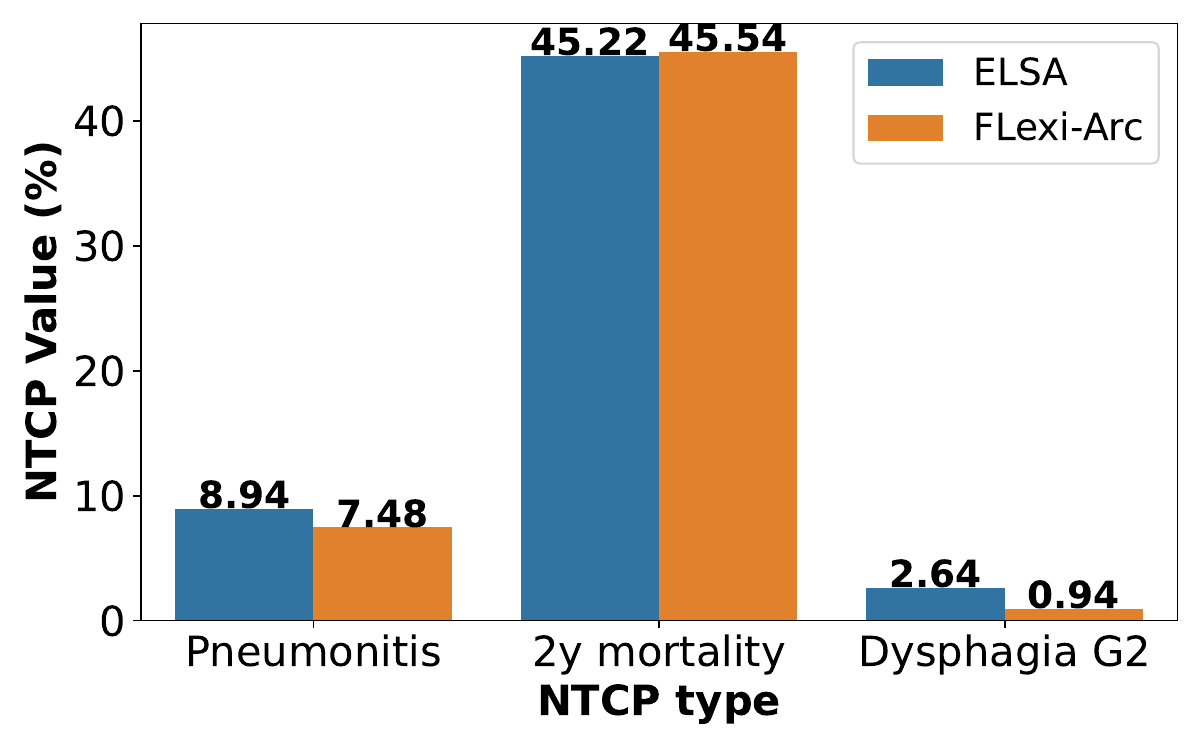}
  \caption{NTCP}
  \label{fig:sub-ntcpLung}
\end{subfigure}
\caption{Comparison of OARs dose and toxicity between the Flexi-Arc and ELSA plans (\textbf{lung} tumor case). Negative differences imply that the Flexi-Arc plan was better than the ELSA plan for that metric (i.e., less dose to the OARs).}
\label{fig:toxicityLung}
\end{figure}

Figure \ref{fig:sub-oarLung} displays the EL pre-selection for both methods and the resulting dose distribution after spot weight optimization. Flexi-Arc did not fully utilize the available half-arc range. In terms of dose distribution, both methods effectively spare the heart, achieving similar results. Figure \ref{fig:sub-ntcpLung} provides a more detailed comparison of OAR metrics, showing that Flexi-Arc achieves better maximum dose sparing of the esophagus compared to ELSA, with the remaining metrics being closely matched.

Finally, NTCP was evaluated for this patient, using the model from a prior study \cite{chocan_dosimetric_2024}. Both methods resulted in comparable outcomes in terms of NTCP, with Flexi-Arc showing a slight improvement in reducing pneumonitis and grade II dysphagia risks. However, the absolute values for these risks were low, indicating minimal overall risk for either method.

\begin{figure}[h!]
    \centering
    \includegraphics[width=0.8\linewidth]{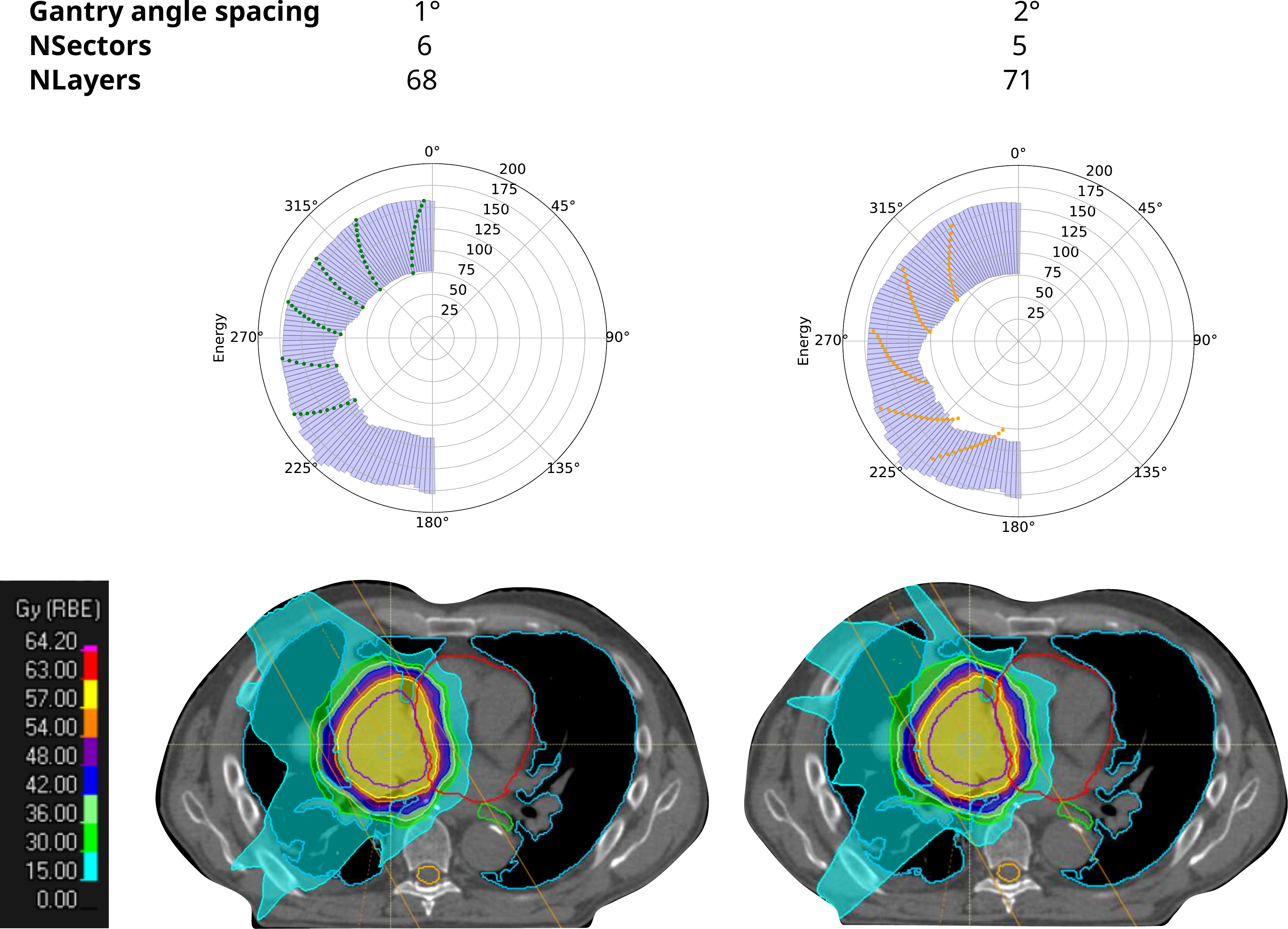}
    \caption{Energy layer pre-selection selected using the Flexi-Arc (left) and ELSA (right) method in the \textbf{lung} tumor case. Final dose distributions are displayed below after spot weight optimization in the RayStation TPS.}
    \label{fig:lungDose}
\end{figure}

\begin{table}[h!]
\centering
\begin{tabular}{llll}
\toprule
                             & \textbf{Flexi-Arc} & \textbf{ELSA}  & \textbf{IMPT}              \\ \midrule
\textbf{Gantry angle spacing (°) }       & 1        & 2    & \multirow{2}{*}{} \\
\textbf{NSectors    }                 & 6         & 5     &                   \\ \midrule
CTV D98\%-nom (Gy{[}RBE{]}) & 58.7      & 58.8  &      \textbf{58.9}             \\
CTV D95\%-wc (Gy{[}RBE{]})   & \textbf{57.7}      & 57.0  &    57.0               \\
CTV D1\%-nom (Gy{[}RBE{]})   & 61.4      & \textbf{61.2}  &    \textbf{61.2}               \\
CI                           & \textbf{2.12}      & 2.30  &   2.90                \\
HI                           & 0.05      & \textbf{0.04}  &   \textbf{0.04}                \\
Patient ID ( Gy $\cdot$ L)   & \textbf{90.7}      & 98.1  &   100.0                \\ \midrule
NSpots                       & 9527      & 9323 & 4816              \\
NLayers                      & 68        & 71   & 29                \\
\textit{Ideal} BDT (s)      & \textbf{102.0}      & 122.5 & /   \\ \bottomrule             
\end{tabular}
\caption{Dosimetric evaluation and plan metrics for the \textbf{lung} case.}
\label{tab:lungDosi}
\end{table}

\section{Discussion}
In this study, we introduced a new method, Flexi-Arc, for pre-selecting energy layers in the context of PAT treatment planning optimization. To validate this method, we applied it to three treatment sites and generated robust PAT treatment plans. We compared these plans with the state-of-the-art ELSA pre-selection method by RaySearch and conventional IMPT plans.

The results demonstrate that the Flexi-Arc method generated EL pre-selections with a good robust target coverage and a short BDT for all three considered patients and this within a very short computation time. A smaller amount of energy layers were sufficient to meet clinical goals. For the brain tumor case, Flexi-Arc plans outperformed ELSA plans with 90 and 120 energy layers due to better sector and EL pre-selection. Their quality was similar to that of the ELSA plan with 180 energy layers, while achieving a 30\% reduction in BDT with the 16-sector plan. The results for the esophagus and lung tumor cases further validated the efficiency of the method, with BDT reductions of 41\% and 17\%, respectively, compared to ELSA plans, and a similar or slightly superior plan quality, depending on the specific metrics evaluated.

A major advantage of the Flexi-Arc technique lies in the freedom it offers for sector placement within the arc. Through meta-heuristics, it optimizes the start and stop positions of each sector to minimize BDT, maximize target coverage, and improve OAR sparing. The sector optimizer will then take benefit of `dead angles' or passage above an OAR without irradiation to make the SU, reducing the need to brake and, hence ensuring smooth gantry rotation velocity. The method allows for a less sector-dependent approach with BDT being more correlated with the total number of energy layers used. It also enables planners to tailor treatment plans to each patient's specific needs, in particular by prioritizing OARs to spare. In contrast, ELSA's sector spacing is constrained by user-defined gantry angle spacing. As a result, it may require gantry slowdown or null velocity if spacing is too small to give time to carry out the SU before the next sector begins, thus prolonging the BDT. ELSA's BDT is therefore highly dependent on the pre-determined number of sectors, which, in turn, relies on user-defined arc beam range and angle spacing. In this study, ELSA plans were initialized with a user-defined arc range and gantry angle spacing. The plans could have been initialized differently based on the planner's intuition but the configuration choice is not straightforward. An experienced planner would have most probably tried to restrict the arc beam range such that it avoids some critical structures. For example, they could have used two partial arc beams with finer spacing for the brain tumor case. Nevertheless, Flexi-Arc could be useful to compute the best arc directions and to suggest it to the planner for consideration. This demonstrates that the planning process depends not only on the method but also on the planner's experience and strategies in employing the available degrees of freedom.

Another advantage, as well as a potential bottleneck, in the proposed EL pre-selection method is the use of evolutionary algorithms with parallel computation. These algorithms do not involve derivatives, making them suitable for complex problems as the ones we are dealing with here. This approach offers the advantage of fast optimization and the exploration of the extensive parameter space for EL pre-selection. It enables us to assess EL pre-selection with varying numbers of sectors and refine the Pareto front for specific sector number by employing different starting points and algorithms. It then ensures a diverse initial population and a wide covering of objectives values. Yet, it is noteworthy that these meta-heuristics do not guarantee optimal solutions and require careful parameter tuning. While we consider the non-dominated front as the Pareto front for our analysis, we cannot guarantee that it is indeed the true Pareto front. Points do converge to it, but the degree of convergence may vary depending on factors such as the number of generations, individuals, and other algorithm-specific parameters. Running the problem with different settings might lead to further convergence and refinement of the Pareto front.

Despite its advantages, the Flexi-Arc method introduces additional complexity due to numerous parameters in the EL pre-selection phase. These include parameters for evolutionary algorithms mentioned above, weighting of sub-objective terms, OAR-crossing penalty weights, WET spacing, arc range, and gantry spacing. The choice of OARs for the pre-optimization step is also important. Incorporating too many OARs can become overly difficult for the optimizer as it does not have much freedom left to place the sectors without being penalized. We therefore encourage to prioritize OARs that the planner wishes to avoid at all cost. While this complexity allows for diverse exploration and potential improvements, it also poses a challenge for planners. The decision-making process requires thoughtful parameter tuning to achieve desired outcomes.

Another point of discussion is considering whether the objective terms used in our study, as surrogates for dynamic BDT and plan quality, are suitable metrics. While there is a trend suggesting a correlation between geometrical target coverage and plan quality, the reflected dosimetric differences between two data points are much smaller than the large geometrical target coverage difference that could be observed in the Pareto plot. According to the results, the dosimetric quality of plans appears to be relate more to the number of energy layers and sectors than to geometric target coverage directly, which may result from the use of a higher number of ELs or sectors. Therefore, one should exercise caution when selecting the EL pre-selection with the highest minimal target coverage, as it may not necessarily translate to the highest plan quality. In the future, our BDT proxy could be improved by considering factors like gantry acceleration. However, since we do not know the number of spots per layer in the pre-selection stage, we will still be limited in modeling the actual expected layer delivery time.

Additional improvement of the Flexi-Arc method can also be made. For instance, refining the WET sequence generation within sectors could optimize coverage further instead of using the simple linear WET spacing between the sector start and end. Incorporating robustness parameters directly into the objective function during pre-selection optimization is another avenue for enhancement. Finally, applying this method to a broader range of patient cases and treatment sites will help demonstrate its potential,versatility, and limitations.

\section{Conclusion}

In conclusion, the Flexi-Arc method introduces an efficient approach to EL pre-selection in PAT treatment planning. This pre-selection involves optimizing the angular position of decreasing energy levels within a specified arc range. The solutions are evaluated using a fast bi-objective function, which combines a metric for geometrical target coverage and a direct estimate of dynamic beam delivery time. This evaluation is performed through multi-objective evolutionary algorithms. The resulting solutions form a Pareto front, allowing the planner to choose the best trade-offs based on the clinical context. Flexi-Arc EL pre-selection achieves high plan quality after dose optimization and expedites delivery within shorter times than other state-of-the-art EL pre-selection methods. Flexi-Arc achieves high plan quality after dose optimization and significantly reduces delivery time compared to other state-of-the-art EL pre-selection methods. However, it requires complex heuristics and parameter tuning to maintain efficiency without sacrificing efficacy. This highlights the need for careful crafting of such heuristics to ensure they are well-aligned with the problem's complexity. Despite these challenges, Flexi-Arc’s flexibility, parallel computation, and high potential for subsequent in spot weight optimization make it a promising avenue for PAT treatment planning.

\section{Acknowledgments}
Sophie Wuyckens is funded by the Walloon Region as part of the Arc Proton Therapy convention (Pôles Mecatech et Biowin). Computational resources have been provided by the supercomputing facilities of the Université catholique de Louvain (CISM/UCL) and the Consortium des Équipements de Calcul Intensif en Fédération Wallonie Bruxelles (CÉCI) funded by the F.R.S.-FNRS under convention 2.5020.11. John A.~Lee is a Research Director with the F.R.S.-FNRS. We would like to thank RaySearch Laboratories AB (Stockholm, Sweden) who granted us access to a modified version built of RayStation 2023B to conduct the study.

\bibliography{main}
\bibliographystyle{unsrt}

\clearpage
\section*{Supplementary Material}
    
\beginsupplement

\section*{SM 1. Beam geometry for planning}
\begin{table}[h!]
\centering
\begin{tabular}{llll}
\toprule
                         & Brain        & Esophagus  & Lung          \\ \midrule
\multicolumn{4}{l}{\textbf{IMPT}}                                             \\ 
Beams (°)                & (70,70,315)  & (150,180)  & (190,240,290) \\
Couch kick (°)           & (220,320,90) & (0,0)      & (0,0,0)       \\ \midrule
\multicolumn{4}{l}{\textbf{ELSA}}                                             \\ 
Arc range (°)            & 0-360 (CW)   & 0-360 (CW) & 190-330 (CCW)   \\
Gantry angle spacing (°) & 2,3,4        & 2          & 2             \\ \midrule
\multicolumn{4}{l}{\textbf{Flexi-Arc}}                                        \\ 
Arc range (°)            & 0-360 (CW)   & 0-360 (CW) & 180-360 (CCW)   \\
Gantry angle spacing (°) & 1            & 1          & 1      \\ \bottomrule      
\end{tabular}
\caption{Beam configuration for the three different planning methods for each treatment site investigated in the study.}
\label{tab:beamConfig}
\end{table}


\section*{SM 2. Water equivalent thickness map}
\begin{figure}[h!]
    \centering
    \includegraphics[width=0.8\linewidth]{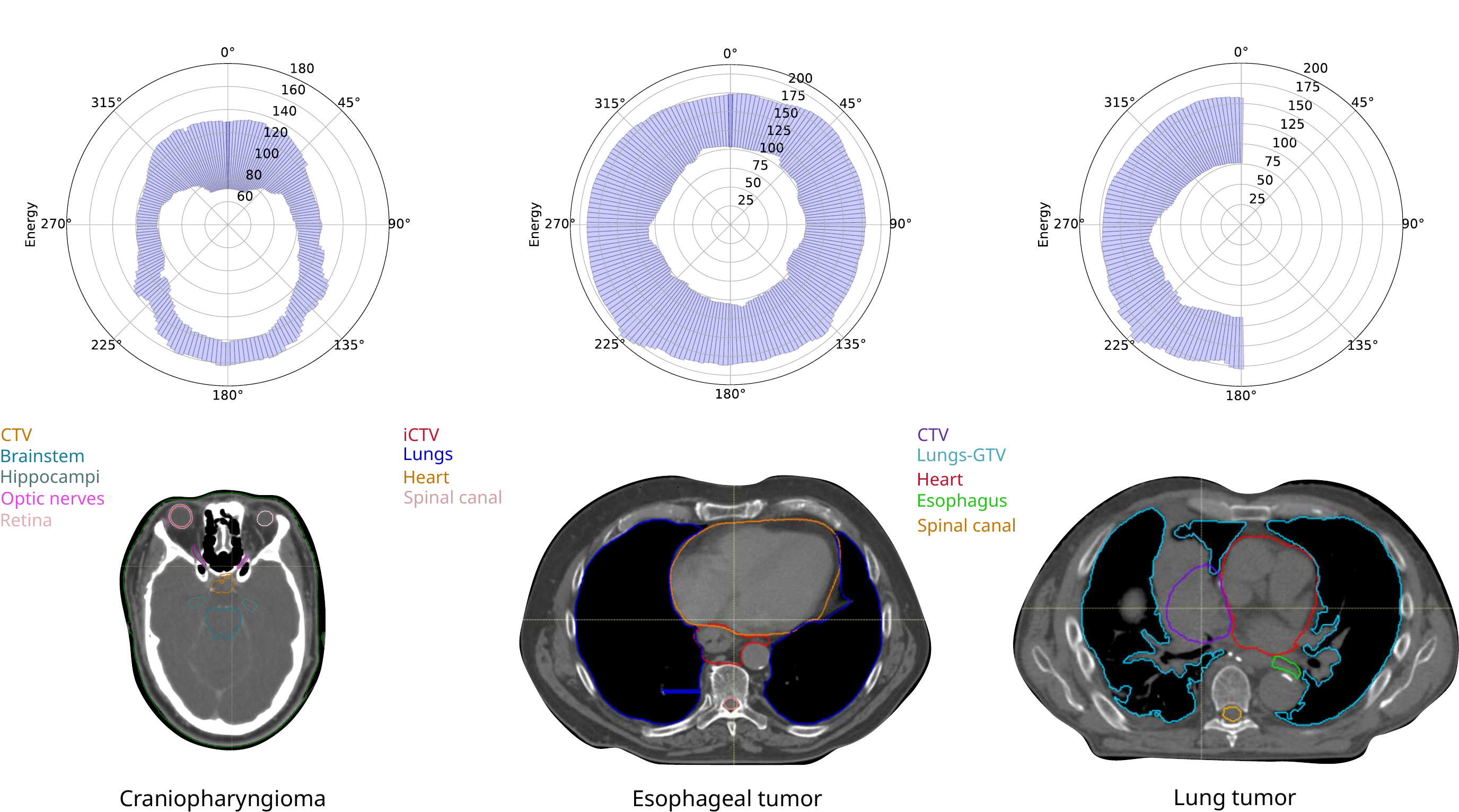}
    \caption{Patients geometry displayed with pre-computed WET ranges (converted to energy ranges [MeV]).}
    \label{fig:brainGeom}
\end{figure}

\section*{SM 3. Clinical goals}

\subsection*{A. Craniopharyngioma}
Table \ref{tab:clinicGoals} lists the clinical goals in the field of neuro-oncology (adapted from Lambrecht et al, 2018, Radiotherapy and Oncology) we referred to for the planning. Doses were recalculated to equivalent dose in 1.8~Gy-fractions using the equivalent dose in 2~Gy-fractions (EQD2) formula $\text{EQD2} = \frac{n \cdot d (d + \alpha/\beta)}{2 + \alpha/\beta}$ where $n = 30$ is the number of fractions used for the treatment and $d$ is the unknown computed for each dose constraint.

\begin{table}[h!]
\centering
{\fontsize{10}{12}\selectfont
\begin{tabular}{ccc}
\toprule
 \textbf{Region of interest (ROI)} & \textbf{$\alpha/\beta$ (Gy)} & \textbf{Clinical goal} \\ \midrule
 CTV (nominal) & / & D98\%   \textgreater 52.92~Gy (98\% Dp = 52.92~Gy) \\
 CTV (worst-case) & / &  D98\%   \textgreater 51.3~Gy (95\%Dp = 51.3~Gy) \\\midrule 
 Brain & 2 &   V60Gy $\le$ 3~cc \\
 \multirow{2}{*}{Brainstem} & \multirow{2}{*}{2}    &  Surface: D0.03~cc $\le$ 60~Gy \\
& & Interior: D0.03~cc $\le$ 56~Gy \\
 Chiasm \& Optic nerve  & 2 &  D0.03~cc $\le$ 56.5~Gy \\
 Cochlea   & 3 &  Dmean $\le$ 48.5~Gy \\
 Hippocampus & 2 &  D40\%   $\le$ 12~Gy \\
 Lacrimal gland & 3 &  Dmean $\le$ 31~Gy \\
 Eyes & 2 & Dmean $\le$ 20~Gy\\
 Lens   & 1 &  D0.03~cc $\le$ 18.5~Gy\\
Retina    & 3 & D0.03~cc $\le$ 48.7~Gy\\ \bottomrule
\end{tabular}%
}
\caption{Clinical goals in neuro-oncology adapted from Lambrecht et al (2018, Radiotherapy and Oncology) for the current fractionation scheme ($30\times1.8$~Gy~(RBE)). Dp = dose prescription, D0.03 cc = near maximum dose to 0.3 cc of structure/organ, Dmean = mean dose, Dx\% = dose to x\% of the considered volume. }
\label{tab:clinicGoals}
\end{table} 

\subsection*{B. Esophageal tumor}

\begin{table}[h!]
\centering
\begin{tabular}{lll}
\toprule
\textbf{ROI}          & \textbf{Clinical goal}                                                                                    & \textbf{Worst case} \\ \midrule
iCTV         & \begin{tabular}[c]{@{}l@{}}98~\% Dp $\le$ Dmean $\le$ 102~\% Dp\\ V95~\% $\ge$ 99~\%\end{tabular}           & V95~\% $\ge$ 97~\%     \\ \midrule
Spinal canal & D0.05~cc $\le$ 45~ Gy                                                                           & D0.05~cc $\le$ 50~ Gy  \\
Body & \begin{tabular}[c]{@{}l@{}}D0.05~cc $\le$ 110~\%\\ D1~cc $\le$ 107~\%\end{tabular} & \begin{tabular}[c]{@{}l@{}}D1~cc $\le$ 110~\%\\ D5~cc $\le$ 107~\%\end{tabular} \\
Lungs        & \begin{tabular}[c]{@{}l@{}}MLD $\le$ 20~ Gy\\ V20~ Gy $\le$ 35~\%\\ V5~ Gy $\le$ 70~\%\end{tabular}  &                       \\
Heart        & \begin{tabular}[c]{@{}l@{}}MHD $\le$ 26~ Gy\\ V40~ Gy $\le$ 30~\%\\ V25~ Gy $\le$ 50~\%\end{tabular} &                \\ \bottomrule     
\end{tabular}
\caption{Clinical goals for target and OARs in esophagus planning. Heart clinical goals are secondary objectives. Some evaluation criteria are also part of the robustness evaluation. Dp = dose prescription (50.4~Gy), MHD = Mean Heart Dose, MLD = Mean Lung Dose.}
\label{tab:esoGoals}
\end{table}

\newpage

\subsection*{C. Lung tumor}

\begin{table}[h!]
\centering
\begin{tabular}{lll}
\toprule
\textbf{ROI}          & \textbf{Clinical goal}                                                                                    & \textbf{Worst case}          \\ \midrule
CTV & \begin{tabular}[c]{@{}l@{}}D98~\% $\geq$ 57~Gy\\ D1~\% $\le$ 63~Gy\end{tabular} & D95~\% $\geq$ 57~Gy \\ \midrule
Spinal canal & D0.04~cc $\le$ 50~Gy                                                                  & D0.04~cc $\le$ 50~Gy \\
Esophagus    & D0.04~cc $\le$ 60~Gy                                                                  &                                \\
Body         & D1~cc $\le$ 63~Gy                                                        & D1~cc $\le$ 63~Gy    \\
Lungs - GTV  & \begin{tabular}[c]{@{}l@{}}Dmean $\le$ 20 ~Gy\\ V30~Gy $\le$ 20~\%\end{tabular}  &                                \\
Heart        & \begin{tabular}[c]{@{}l@{}}Dmean $\le$ 20~Gy\\ D0.04~cc $\le$ 60~Gy\end{tabular} &             \\ \bottomrule                  
\end{tabular}
\caption{Clinical goals for target and OARs in lung planning. }
\label{tab:lungGoals}
\end{table}

\newpage

\section*{SM 4. Pareto fronts}

\newpage
\section*{SM 5. IMPT dose distributions}

\begin{figure}[h!]
    \centering
    \includegraphics[width=0.8\linewidth]{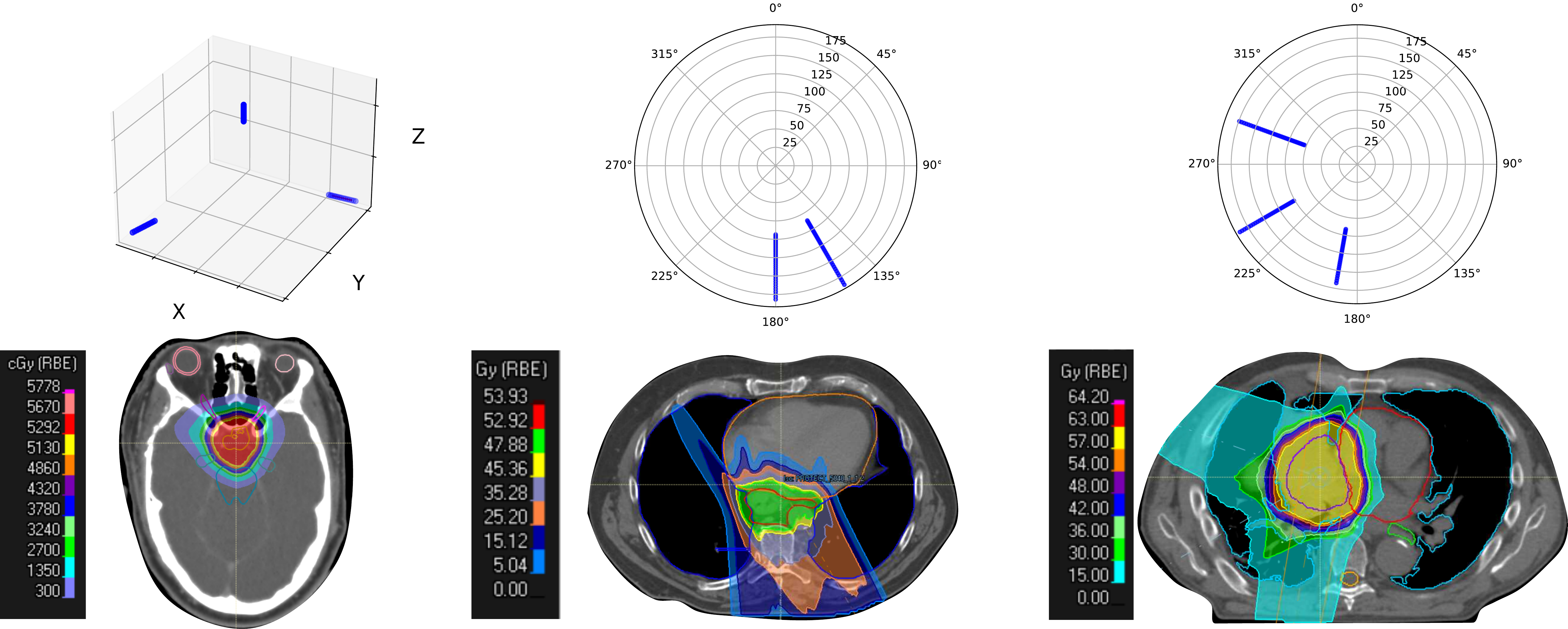}
    \caption{Energy layers for each IMPT beam and their associated dose distribution for the three treatment sites. Craniopharyngioma (left), esophageal tumor (center) and lung tumor (right).}
    \label{fig:enter-label}
\end{figure}

\section*{SM 6. Organs-at-risk dosimetry}
\subsection*{A. Craniopharyngioma}

\begin{table}[h!]
\centering
{\fontsize{10}{12}\selectfont
\begin{tabular}{@{}lcccccc@{}}
\toprule
\textbf{} & \multicolumn{2}{c}{\textbf{Flexi-Arc}} & \multicolumn{3}{c}{\textbf{ELSA}} & \textbf{IMPT} \\ 
\cmidrule(l){2-3} \cmidrule(l){4-6} 
 \textbf{Gantry angle spacing ($^{\circ}$)} & 1 & 1 & 4 & 3 & 2 & \\
 \textbf{NSectors}         & 8   & 16 & 5 & 5 & 8  & \\ 
\midrule 
D0.03cc Brain stem interior (Gy)    &  51.1 & 49.4 &  52.0 & 53.3& 51.5 & 51.1\\
D0.03cc Brain stem surface (Gy)      & 54.5 & 53.2&   54.3& 54.9& 54.7 & 54.2\\
D0.03cc Chiasm (Gy)  & 55.3 & 54.8 &   55.2& 55.6& 55.3 & 55.5\\
D0.03cc Optic nerve L (Gy)  & 53.5 & 52.9 &   52.9& 52.1& 52.4 & 55.0\\
D0.03cc Optic nerve R (Gy) & 50.9 & 51.0 &   53.5& 49.5& 50.3 & 53.1\\
D0.03cc Lens L (Gy) & 0.7 & 0.2 &    0.5& 0.3&0.7& 0.0\\
D0.03cc Lens R (Gy)     & 0.4 & 0.2 &    0.9&1.1 &0.3& 0.0\\
D0.03cc Retina L (Gy)     & 3.0 & 1.6 &   3.4 & 2.9&2.9& 0.0\\
D0.03cc Retina R (Gy) & 1.2 & 0.9 &  3.1 & 2.5& 1.1 & 0.0 \\
D40\% Hippocampus L (Gy)  & 1.0 & 0.9 &  3.5 & 3.3&1.1 & 0.2\\
D40\% Hippocampus R (Gy)  & 2.1 & 2.3 &  4.0 &3.2 & 2.6 & 0.6\\
Dmean Eye L (Gy)  & 0.9  & 0.6 &   1.0 &1.0 & 1.0 & 0.0\\
Dmean Eye R (Gy)  & 0.5 & 0.3 &  1.0 &  1.0& 0.5 & 0.0\\
Dmean Lacrimal gland L (Gy)  & 0.1 & 1.1 & 1.5 & 2.4 & 1.0 & 0.0 \\
Dmean Lacrimal gland R (Gy)  & 1.0 & 0.6 & 0.6 & 0.6 & 1.0 & 0.0\\
Dmean Cochlea L (Gy)  & 1.0 & 0.7 &  2.0& 2.0 & 1.0 & 0.0\\
Dmean Cochlea R (Gy)  & 0.9 &0.8  & 2.0  &2.0  & 0.9 & 0.0\\
\bottomrule
\end{tabular}%
}
\caption{Detailed dosimetric comparison for organs-at-risk in the \textbf{nominal} scenario (\textbf{brain} case). \textit{Abbreviations :  D0.03 cc = near maximum dose to 0.3 cc of structure/organ, Dmean = mean dose, D40\% = dose to 40\% of the volume of both hippocampi, L = left, R = right.}}
\label{tab:table1_p1}
\end{table}

\begin{table}[h!]
\centering
{\fontsize{10}{12}\selectfont
\begin{tabular}{@{}lcccccc@{}}
\toprule
\textbf{} & \multicolumn{2}{c}{\textbf{Flexi-Arc}} & \multicolumn{3}{c}{\textbf{ELSA}} & \textbf{IMPT} \\ 
\cmidrule(l){2-3} \cmidrule(l){4-6} 
 \textbf{Gantry angle spacing ($^{\circ}$)} & 1 & 1 & 4 & 3 & 2 & \\
 \textbf{NSectors}         & 8   & 16 & 5 & 5 & 8  & \\ 
\midrule 
D0.03cc Brain stem interior (Gy)    & 53.8 &52.4 &  53.4 & 54.6& 54.2 & 53.9\\
D0.03cc Brain stem surface (Gy)      & 55.7 &54.3 &   55.2& 55.8 & 55.7 & 56.0\\
D0.03cc Chiasm (Gy)  & 55.8 & 55.7 &   55.9& 56.3& 55.9 & 55.8\\
D0.03cc Optic nerve L (Gy)  & 54.8 & 55.3 &   55.2&54.4 & 53.7 & 55.5\\
D0.03cc Optic nerve R (Gy) & 54.0 & 53.3 &   56.5&52.8 & 54.0 & 54.5\\
D0.03cc Lens L (Gy) &  1.0& 0.3 &   0.9& 0.7& 1.0 & 0.0\\
D0.03cc Lens R (Gy)     & 0.0 & 0.3 &   1.2& 1.2& 0.5 & 0.0 \\
D0.03cc Retina L (Gy)     & 5.3 & 2.5 &   4.2& 4.9& 3.8& 0.0\\
D0.03cc Retina R (Gy) & 2.3 & 1.8 &   4.2& 3.3& 1.5 & 0.0\\
D40\% Hippocampus L (Gy)  & 1.6 & 1.5 &   4.7& 4.8 & 1.8 & 0.6 \\
D40\% Hippocampus R (Gy)  & 3.5 & 3.3 &  5.3& 4.5 & 3.9 & 1.3\\
Dmean Eye L (Gy)  & 1.2 & 0.8 &  1.5 & 1.6 & 1.4 & 0.0\\
Dmean Eye R (Gy)  & 0.7 &  0.4 & 1.3 & 1.2 & 0.6 & 0.0\\
Dmean Lacrimal gland L (Gy)  & 0.2 & 1.5 & 2.3  & 3.0 & 1.2  & 0.0 \\
Dmean Lacrimal gland R (Gy)  & 1.4 & 0.9 & 0.7 & 0.9 & 1.4& 0.0\\
Dmean Cochlea L (Gy)  & 1.6 & 1.3  & 2.8 & 2.9 & 1.9 & 0.0\\
Dmean Cochlea R (Gy)  & 1.9 & 1.5 & 2.8 & 3.5 & 1.6 & 0.0 \\
\bottomrule
\end{tabular}%
}
\caption{ Detailed dosimetric comparison for organs-at-risk in the \textbf{worst-case} scenario (\textbf{brain} case). \textit{Abbreviations :  D0.03 cc = near maximum dose to 0.3 cc of structure/organ, Dmean = mean dose, D40\% = dose to 40\% of the volume of both hippocampi, L = left, R = right.}}
\label{tab:table1_p1_wc}
\end{table}
\newpage

\subsection*{B. Esophageal tumor}

\begin{table}[h!]
\centering
\begin{tabular}{llllllll} \toprule
 & & \multicolumn{3}{l}{Nominal} & \multicolumn{3}{l}{Worst-case} \\ \midrule
                       &                  & Flexi-Arc & ELSA & IMPT & Flexi-Arc & ELSA & IMPT \\ \cmidrule(l){3-5}  \cmidrule(l){6-8} 
\multirow{3}{*}{Lungs} & Dmean {[}Gy{]}   & 4.6       & 4.5  & 4.1  & 5.4 & 5.5 & 4.7\\
                       & V20 {[}\%{]}     & 5.1       & 3.8  & 10.0 & 6.8 & 5.2 & 10.9\\
                       & V5 {[}\%{]}      & 28.5      & 28.5 & 15.9 & 33.1 & 35.0 & 19.3\\
\multirow{3}{*}{Heart} & Dmean {[}Gy{]}   & 6.6       & 8.1 & 8.1  & 9.5 & 11.3 & 11.3\\
                       & V40 {[}\%{]}     & 6.7       & 7.1 & 8.3  & 11.0 & 11.5 & 13.0\\
                       & V25 {[}\%{]}     & 10.7      & 11.8 & 13.6 & 16.3 & 17.7 & 20.3\\
Spinal canal           & D0.05cc {[}Gy{]} & 28.7      & 28.7 & 39.6 & 37.6 & 33.6 & 39.9\\
Body                   & D1cc {[}Gy{]}    & 53.0      & 52.7 & 52.7 & 54.0 & 54.5 & 53.5\\ \bottomrule
\end{tabular}
\caption{Detailed dosimetric comparison for organs-at-risk (OAR) for the \textbf{esophagus} case. \textit{Abbreviations :  D0.05 cc = near maximum dose to 0.05 cc of structure/organ, Dmean = mean dose, V20 = volume that receives at least 20 Gy}.}
\label{tab:esoDosiOARs}
\end{table}

\newpage

\subsection*{C. Lung tumor}

\begin{table}[h!]
\centering
\begin{tabular}{llllllll} \toprule
 & & \multicolumn{3}{l}{Nominal} & \multicolumn{3}{l}{Worst-case} \\ \midrule
                       &                  & Flexi-Arc & ELSA & IMPT & Flexi-Arc & ELSA & IMPT \\ \cmidrule(l){3-5}  \cmidrule(l){6-8} 
\multirow{2}{*}{Esophagus} & Dmean {[}Gy{]}   & 1.0       & 2.0  & 1.8  & 4.2       & 4.6  & 4.3  \\
                       & D0.04cc {[}\%{]}     & 12.4       & 25.1  & 26.6 & 38.4       & 37.4  & 42.4 \\
\multirow{2}{*}{Lungs - GTV} & Dmean {[}Gy{]}   & 6.1      & 7.5  & 7.6  & 7.8       & 9.2  & 9.4  \\
                       & V30 {[}\%{]}     & 6.2      & 7.0  & 8.3 & 9.1       & 9.7  & 11.3 \\
\multirow{3}{*}{Heart} & Dmean {[}Gy{]}   & 5.2       & 5.0  & 4.9  & 7.8       & 7.2 & 7.2 \\
                       & D0.04cc {[}\%{]}     & 61.2       & 62.1  & 62.4  & 64.6      & 75.7 & 63.6 \\
Spinal canal           & D0.05cc {[}Gy{]} & 6.2      & 6.9 & 18.8 & 23.2      & 13.9 & 19.2 \\
Body                   & D1cc {[}Gy{]}    & 62.9      & 61.7 & 62.3 & 63.7     & 67.4 & 62.5 \\ \bottomrule
\end{tabular}
\caption{Detailed dosimetric comparison for organs-at-risk (OAR) for the \textbf{lung} case. \textit{Abbreviations :  D0.04 cc = near maximum dose to 0.04 cc of structure/organ, Dmean = mean dose, V30 = volume that receives at least 30 Gy}.}
\label{tab:lungDosiOARs}
\end{table}

\end{document}